# Parsimonious Optimal Dynamic Partial Order Reduction

Parosh Aziz Abdulla[1], Mohamed Faouzi Atig[1], Sarbojit Das[1],
Bengt Jonsson[1], and Konstantinos Sagonas[1,2]

[1] Uppsala University, Uppsala, Sweden
[2] National Technical University of Athens, Athens, Greece

**Abstract.** Stateless model checking is a fully automatic verification technique for concurrent programs that checks for safety violations by exploring all possible thread schedulings. It becomes effective when coupled with Dynamic Partial Order Reduction (DPOR), which introduces an equivalence on schedulings and reduces the amount of needed exploration. DPOR algorithms that are *optimal* are particularly effective in that they guarantee to explore *exactly* one execution from each equivalence class. Unfortunately, existing sequence-based optimal algorithms may in the worst case consume memory that is exponential in the size of the analyzed program. In this paper, we present Parsimonious-OPtimal DPOR (POP), an optimal DPOR algorithm for analyzing multi-threaded programs under sequential consistency, whose space consumption is polynomial in the worst case. POP combines several novel algorithmic techniques, including (i) a parsimonious race reversal strategy, which avoids multiple reversals of the same race, (ii) an eager race reversal strategy to avoid storing initial fragments of to-be-explored executions, and (iii) a space-efficient scheme for preventing redundant exploration, which replaces the use of sleep sets. Our implementation in NIDHUGG shows that these techniques can significantly speed up the analysis of concurrent programs, and do so with low memory consumption. Comparison to TruSt, a related optimal DPOR algorithm that represents executions as graphs, shows that POP's implementation achieves similar performance for smaller benchmarks, and scales much better than TruSt's on programs with long executions.

## 1 Introduction

Testing and verification of multi-threaded programs is challenging, since it requires reasoning about all the ways in which operations executed by different threads can interfere. A successful technique for finding concurrency bugs in multithreaded programs and for verifying their absence is *stateless model checking* (SMC) [19]. Given a terminating program and fixed input data, SMC systematically explores the set of all thread schedulings that are possible during program runs. A dedicated runtime scheduler drives the SMC exploration by making decisions on scheduling whenever such choices may affect the interaction between threads. Given enough time, the exploration covers all possible executions and detects any unexpected program results, program crashes, or assertion violations. The technique is entirely automatic, has no false positives, does not consume excessive memory, and can reproduce the concurrency bugs it detects. SMC has been implemented in many tools (e.g., VeriSoft [20], CHESS [39], Concuerror [14], NIDHUGG [2], rInspect [48], CDSCHECKER [41], RCMC [28], and GENMC [34]), and successfully applied to realistic programs (e.g., [21] and [32]).

To reduce the number of explored executions, SMC tools typically employ *dynamic partial order reduction* (DPOR) [1,17,28]. DPOR defines an equivalence relation on executions, typically Mazurkiewicz trace equivalence [36], which preserves many important correctness properties, such as reachability of local states and assertion violations, and explores at least one execution in each equivalence class. Thus, to analyze a program, it suffices to explore one execution from each equivalence class. DPOR was originally developed [17] for models of concurrency where executions are expressed as sequences of interactions between threads/processes and shared objects. Subsequently, sequence-based DPOR has been adapted and refined to a number of programming models, including actor programs [46], abstract computational models [27], event driven programs [4,24,35], and MPI programs [42]; it has been extended with features for efficiently handling spinloops and blocking constructs [25], and been adapted for weak concurrency memory models, such as TSO and PSO [2,48]. DPOR has also been adapted for weak memory models by representing executions as graphs, where nodes represent read and write operations, and edges represent reads-from and coherence relations; this allows the algorithm to be parametric on a specific memory model, at the cost of calling a memory-model oracle [28,30]

An important improvement has been the introduction of *optimal* DPOR algorithms, which are efficient in that they guarantee to explore *exactly* one execution from each equivalence class. The first optimal DPOR algorithm was designed for the sequence-based representation [1]. Subsequently, optimal DPOR algorithms for even weaker equivalences than Mazurkiewicz trace equivalence have been developed [5,8,10]. In some DPOR algorithms [1,8,10], optimality comes at the price of added memory consumption which in the worst case can be exponential in the size of the program [3]. Even though most benchmarks in the literature show a modest memory overhead as the price for optimality, it would be desirable to have an optimal DPOR algorithm whose memory consumption is guaranteed to be polynomial in the size of the program. Such an algorithm, called TruSt [29], was recently presented, but for a graph-based setting [30]. It would be desirable to develop a polynomial-space optimal DPOR algorithm also for sequence-based settings. One reason is that a majority of past work on DPOR is sequence-based; hence such an algorithm could be adapted to various programming models and features, some of which were recalled above. Another reason is that sequence-based models represent computations adhering to sequential consistency (SC) and TSO more naturally than graph-based models. For SC, representing executions as sequences of events makes executions consistent by construction and alleviates the need to resort to a potentially expensive memory-model oracle for SC.

In this paper, we present the Parsimonious-OPtimal DPOR (POP) algorithm for analyzing multi-threaded programs under SC (Section 4). POP is designed for programs in which threads interact by atomic reads, writes, and RMWs to shared variables, and combines several novel algorithmic techniques.

- A *parsimonious race reversal* technique (Section 4.1), which considers a race if and only if its reversal will generate a previously unexplored execution; in contrast, most existing DPOR algorithms reverse races indiscriminately, only to thereafter discard redundant reversals (e.g., by sleep sets or similar mechanisms).
- An *eager race reversal* strategy (Section 4.2), which immediately starts exploration of the new execution resulting from a race reversal; this prevents accumulation of a potentially exponential number of execution fragments generated by race reversals.
- In order to avoid exploring several executions in the same equivalence class, a naïve realization of POP would employ an adaptation of sleep sets [18]. However, these can in the worst case become exponentially large. Therefore, POP employs a *parsimonious characterization* of sleep sets (Section 4.3): instead of representing the elements of the sleep set explicitly, POP uses a characterization of them, which allows to detect and prevent redundant exploration, and uses at most polynomial space. This sleep set characterization is computed only from its generating race, implying that explorations of different executions share no state, making POP suitable for parallelization.

We prove that the POP algorithm is *correct* (explores at least one execution in each equivalence class), *optimal* (explores exactly one execution in each equivalence class), does not suffer from blocked explorations, and requires only polynomial size memory.

We have implemented POP DPOR in an extension of the Nidhugg tool [2]. Using a wide variety of benchmarks (Section 6), which are available in the paper's artifact, we show that POP's implementation indeed has its claimed properties, it always outperforms Optimal DPOR's implementation, and offers performance which is on par with TruSt's, the state-of-the-art graph-based DPOR algorithm. Moreover, by being sequence-based, it scales much better than TruSt's implementation on programs with long executions.

## 2 Main Concepts

In this section, we informally present the core principles of our approach, in particular the three novel algorithmic techniques of parsimonious race reversal, eager race reversal, and parsimonious characterization of sleep sets, along with how they relate to previous sequence-based DPOR algorithms, on a simple example, shown in Fig. 1. In this code, four threads $(p, q, r, s)$ access three shared variables (g, x, y, z), using five thread-local registers $(a, b, c, d, e)$.[3] DPOR algorithms typically first ex-

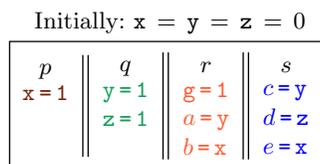

Initially: $x = y = z = 0$

| $p$ | $q$ | $r$ | $s$ |
|---|---|---|---|
| x = 1 | y = 1 | g = 1 | c = y |
|  | z = 1 | a = y | d = z |
|  |  | b = x | e = x |

**Fig. 1.** Program code.

---

[3] Throughout this paper, we assume that threads are spawned by a `main` thread, and that all shared variables get initialized to 0, also by the main thread.

plore an arbitrary execution, which is then inspected to detect races. Assume that this execution is $E_1$ (the leftmost execution in Fig. 2). To detect races in an execution $E$, one first computes its happens-before order, denoted $\xrightarrow{\text{hb}}_E$, which is the smallest transitive relation that orders two events that (i) are in the same thread, or (ii) access a common shared variable and at least one of them is a write. A *race* consists of two events in different threads that are adjacent in the $\xrightarrow{\text{hb}}_E$ order. In execution $E_1$ there are two races on x, two races on y, and one race on z. The two races on y are marked with yellow arrows, as we are going to discuss them now. POP first reverses the race between events y = 1 and $a$ = y. For each race, a DPOR algorithm constructs an initial fragment of an alternative execution, called a *schedule*, which reverses the race and branches off from the explored execution just before the race. POP constructs a minimal schedule consisting of the events that happen before (in the $\xrightarrow{\text{hb}}_{E_1}$ order) the second event followed by the second event of the race, while omitting the first event of the race, resulting in the event sequence $\langle g = 1 \cdot a = y \rangle$, which is inserted as an alternative continuation after x = 1 (the branch to the right of x = 1).

In comparison, early DPOR algorithms, including the "classic" DPOR algorithm by Flanagan and Godefroid [17] and the Source DPOR algorithm of Abdulla et al. [1] construct a schedule consisting of just one event that can initiate an execution which reverses the race ($\langle g = 1 \rangle$ in this case). Storing just one event saves space, but the execution afterwards is uncontrolled and may deviate from the path towards the second racing event $a$ = y, potentially leading to redundant exploration. To avoid redundancy, we need schedules which consist of paths to the second racing event.

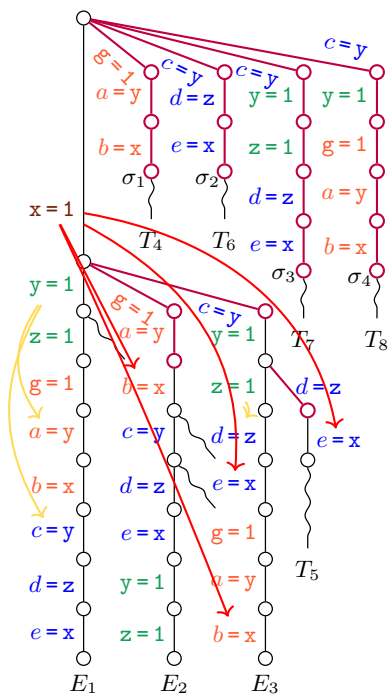

**Fig. 2.** Part of the exploration tree for the program in Fig. 1. Completed executions are denoted $E_i$; truncated subtrees are denoted $T_i$.

*Eager Race Reversal:* Following an eager race reversal strategy, POP continues the exploration with this branch and explores $E_2$. POP can in principle be implemented so that the schedules constructed as alternative continuations of an event are all collected before they are explored. However, such a strategy can in the worst case consume memory that is exponential in the program size. The reason is that, for some programs, the number of schedules that branch off at a particular point in an execution may become exponential in the size of the program; this was first observed by Abdulla et al. [3, Sect. 9]; an illustrating shared-variable program is given by Kokologiannakis *et al.* [29, Sect. 2.3]. POP avoids this problem by *exploring schedules eagerly*: immediately after the creation of a schedule, exploration switches to continuations of that schedule. This strategy can be realized by an algorithm that calls a recursive function to initiate exploration of a new schedule. We establish, in Lemma 1, that the recursion depth of such an algorithm is at most $n(n-1)/2$, where $n$ is the length of the longest execution of the program.

Continuing exploration, POP encounters the race on x in $E_2$ involving events x = 1 and b = x. (In Fig. 2, we show races by red arrows.) POP constructs the schedule $\sigma_1 := \langle g = 1 \cdot a = y \cdot b = x \rangle$ (second branch from the root) and explores the subsequent part of $T_4$ (tree $T_4$ represents all the extensions after $\sigma_1$). After exploring $T_4$ (second branch from the root), POP comes back to $E_2$.

*Parsimonious Race Reversal:* To illustrate POP's mechanism for reversing each race only once, let us next consider races in execution $E_2$. There is one race on y, between $a = y$ and y = 1 in $E_2$, for which POP would construct the schedule $\sigma := \langle c = y \cdot y = 1 \rangle$. However, a prefix of $\sigma$, namely $\langle c = y \rangle$, will be constructed from a race in $E_1$ between y = 1 and c = y and inserted as an alternative continuation after x = 1 (the rightmost child of x = 1 in Fig. 2). Thus, any continuation of $\sigma$ after x = 1 can also be explored as a continuation after the rightmost child of x = 1, implying that inserting $\sigma$ as an alternative continuation after x = 1 would lead to redundant exploration. POP avoids such redundant exploration by forbidding to consider races whose first event (in this case $a = y$) is in some schedule: reversing a race whose first event is in a schedule yields a fragment that is explored in some other execution. The execution $E_2$ also exhibits two races on x, both including x = 1, with the events $b = x$ and $e = x$. These races have already occurred in $E_1$, and should therefore not be considered, since the schedules they would generate will be generated from the corresponding races in $E_1$. POP achieves this by forbidding to consider races whose second event is not fresh. A second event of a race is *fresh* if it happens-after (in the $\xrightarrow{hb}$ order) the last event of each schedule that appears between the two racing events. Returning to the two races on x in $E_2$, their second events are not fresh, and hence they are not reversed.

Let us continue the exploration of $E_2$ in Fig. 2 to illustrate how the eager race reversal strategy affects the order in which branches are explored. In $E_2$, there are two more races, on y and z, whose reversals produce two branches after $b = x$ and $c = y$, denoted by wavy edges. After their exploration, since there are no more races in $E_2$, POP returns to $E_1$, where the race between events y = 1 and c = y induces the schedule $\langle c = y \rangle$, initiating exploration of $E_3$. While exploring $E_3$, the race inolving events z = 1 and $d = z$ in $E_3$ induces the schedule $\langle d = z \rangle$, initiating exploration of the subtree $T_5$, during which the race on x involving x = 1 and $e = x$ induces the schedule $\sigma_2 := \langle c = y \cdot d = z \cdot e = x \rangle$, and explores the subsequent part of the tree $T_6$. After finishing exploration of $T_6$ and $T_5$, POP comes back to $E_3$, where the race involving events x = 1 and $e = x$ induces the schedule $\sigma_3 := \langle c = y \cdot y = 1 \cdot z = 1 \cdot d = z \cdot e = x \rangle$ initiating exploration of $T_7$, whereafter exploration of $E_3$ resumes.

*Parsimonious Characterization of Sleep Sets:* Even though the parsimonious race reversal strategy guarantees that the initial fragments of alternative executions are inequivalent, one must prevent that their continuations become equivalent. This happens when POP continues after a *read schedule*, generated from a race whose second event is a read event. To illustrate this problem, let us consider the race involving events $\mathtt{x\,=\,1}$ and $b\,\mathtt{=\,x}$ in $E_3$, which produces the read schedule $\sigma_4 := \langle c\,\mathtt{=\,y} \cdot \mathtt{y\,=\,1} \cdot \mathtt{g\,=\,1} \cdot a\,\mathtt{=\,y} \cdot b\,\mathtt{=\,x}\rangle$, initiating exploration of $T_8$. Note that the schedule $\sigma_4$ is not conflicting with the read schedules $\sigma_2$ and $\sigma_3$. At this point, we need to be careful: there is a danger that $\sigma_4$ will be continued using the other two schedules ($\sigma_2$ and $\sigma_3$), whereas the explorations starting with schedules $\sigma_2$ and $\sigma_3$ can be continued using $\sigma_4$; we would then explore equivalent executions, consisting of these three schedules in either order. The same problem occurs with $\sigma_1$ and $\sigma_3$, as they do not conflict. The DPOR technique for avoiding such redundant exploration is *sleep sets* [18]. In its standard form, a sleep set is a set of events that should not be performed before some conflicting event. Since POP uses schedules as beginnings of alternative explorations, the appropriate adaptation would be to let a sleep set be a set of read schedules that should not be performed unless some conflicting event is performed before that. In Fig. 2, this would mean that after exploring the continuations of $\sigma_2$ and $\sigma_3$, these schedules are added to the sleep set when starting to explore the continuations of $\sigma_4$, and $\sigma_1$ is added to the sleep set when starting to explore the continuations of $\sigma_3$. This mechanism is simple to combine with parsimonious race reversal and eager exploration of schedules. Unfortunately, there are programs where the number of read schedules that would be added to such a sleep set is exponential in the size of the program, whence the worst-case memory consumption may be exponential in the size of the program. POP avoids this problem by a *parsimonious characterization of sleep sets*, which consumes memory that is polynomial in the size of the program. The idea is to totally order the read schedules. When continuing exploration after a read schedule $\sigma$, the read schedules that precede $\sigma$ in this order are represented by POP's parsimonious characterization in polynomial space, even though the number of represented schedules may be exponential. In principle, there are several ways to order the read schedules. POP uses one such ordering, namely $\sigma_1$, $\sigma_2$, $\sigma_3$ and $\sigma_4$. We provide the details about this representation in Section 4.3.

## 3  Programs, Executions, and Equivalence

We consider programs consisting of a finite set of *threads* that share a finite set of *(shared) variables*. Each thread has a finite set of local registers and runs a deterministic code, built in a standard way from expressions (over local registers) and atomic commands, using standard control flow constructs (sequential composition, selection, and bounded loop constructs). Atomic commands either write the value of an expression to a shared variable, or assign the value of a shared variable to a register, or can atomically both read and modify a shared variable. Conditional control flow constructs can branch on the value of an expression. From here on, we use $t$ to range over threads, and $x, y, z$ to range over shared variables. The local state of a thread is defined as usual by its program counter and the contents of its registers. The global state of a program consists of the local state of each thread together with the valuation of the shared variables. The program has a unique initial state, in which shared variables have predefined initial values. We assume that memory is sequentially consistent.

The execution of a program statement is an *event*, which affects or is affected by the global state of the program. An event is represented by a tuple $\langle t, i, \mathtt{T}, x \rangle$, where $t$ is the thread performing the event, $i$ is a positive integer, denoting that the event results from the $i$-th execution step in thread $t$. $\mathtt{T}$ is the type of the event (either $\mathtt{R}$ for read or $\mathtt{W}$ for write and read-modify-write), and $x$ is the accessed variable. If $e$ is the event $\langle t, i, \mathtt{T}, x \rangle$, we write $e.th$ for $t$, $e.\mathtt{T}$ for $\mathtt{T}$, and $e.var$ for $x$. An *access* is a pair $\langle \mathtt{T}, x \rangle$ consisting of a type and a variable. We write $e.acc$ for $\langle e.\mathtt{T}, e.var \rangle$. We say that two accesses $\langle \mathtt{T}, x \rangle$ and $\langle \mathtt{T}', x' \rangle$ are *dependent*, denoted $\langle \mathtt{T}, x \rangle \bowtie \langle \mathtt{T}', x' \rangle$, if $x = x'$ and at least one of $\mathtt{T}$ and $\mathtt{T}'$ is $\mathtt{W}$. We say that two events $e$ and $e'$ are *dependent*, denoted $e \bowtie e'$, if $e.th = e'.th$ or $e.acc \bowtie e'.acc$. As is customary in DPOR algorithms, we can let an event represent the combined effect of a sequence of statements, if at most one of them accesses a shared variable.

An *execution sequence* (or just *execution*) $E$ is a finite sequence of events, starting from the initial state of the program. We let $\mathtt{enabled}(E)$ denote the set of events that can be performed in the state to which $E$ leads. An execution $E$ is *maximal* if $\mathtt{enabled}(E) = \emptyset$. We let $\mathtt{dom}(E)$ denote the set of events in $E$; we also write $e \in E$ to denote $e \in \mathtt{dom}(E)$. We use $u$ and $w$, possibly with superscripts, to range over sequences of events (not necessarily starting from the initial state), $\langle \rangle$ to denote the empty sequence, and $\langle e \rangle$ to denote the sequence with only the event $e$. We let $w \cdot w'$ denote the concatenation of sequences $w$ and $w'$, and let $w \backslash e$ denote the sequence $w$ with the first occurrence of $e$ (if any) removed. For a sequence $u = e_1 \cdot e_2 \cdot \ldots \cdot e_m$, we let $w \backslash u$ denote $(\cdots ((w \backslash e_1) \backslash e_2) \backslash \cdots) \backslash e_m$.

The basis for a DPOR algorithm is an equivalence relation on the set of execution sequences. The definition of this equivalence is based on a happens-before relation on the events of each execution sequence, which captures the data and control dependencies that must be respected by any equivalent execution.

**Definition 1 (Happens-before).** *Given an execution sequence $E$, we define the* happens-before relation *on $E$, denoted $\xrightarrow{\mathtt{hb}}_E$, as the smallest irreflexive partial order on $\mathtt{dom}(E)$ such that $e \xrightarrow{\mathtt{hb}}_E e'$ if $e$ occurs before $e'$ in $E$, and $e \bowtie e'$.*

The $\mathtt{hb}$-*trace* (or *trace* for short) of $E$ is the directed graph $(\mathtt{dom}(E), \xrightarrow{\mathtt{hb}}_E)$.

**Definition 2 (Equivalence).** *Two execution sequences $E$ and $E'$ are* equivalent, *denoted $E \simeq E'$, if they have the same $\mathtt{hb}$-trace. We let $[E]_\simeq$ denote the equivalence class of $E$.*

The equivalence relation $\simeq$ partitions the set of execution sequences into equivalence classes, paving the way for an optimal DPOR algorithm which explores precisely one execution in each equivalence class.

## 4   Design of the POP Algorithm

In this section, we explain the design of POP, which is optimal in the sense that it explores precisely one execution in each equivalence class defined by Definition 2. We first need some auxiliary definitions

**Definition 3 (Compatible sequences and happens-before prefix).** *For two execution sequences $E \cdot w$ and $E \cdot w'$,*

- *the sequences $w$ and $w'$ are* compatible, *denoted $w \sim w'$, iff there are sequences $w''$ and $w'''$ s.t. $E \cdot w \cdot w'' \simeq E \cdot w' \cdot w'''$,*
- *the sequence $w$ is a* happens-before prefix *of $w'$, denoted $w \sqsubseteq w'$, iff there is a sequence $w''$ s.t. $E \cdot w \cdot w'' \simeq E \cdot w'$.*

We illustrate the definition on the example in Fig. 2. Assuming $E_3 = \langle \mathtt{x\,=\,1} \rangle \cdot w'$, it is true that $\sigma_4 \sqsubseteq w'$, since $\langle \mathtt{x\,=\,1} \rangle \cdot \sigma_4 \cdot w'' \simeq \langle \mathtt{x\,=\,1} \rangle \cdot w'$, where $w''$ is the sequence $\langle \mathtt{z\,=\,1} \cdot d\mathtt{=z} \cdot e\mathtt{=x} \rangle$. However, $\sigma_1 \not\sim \sigma_4$, since $\sigma_1$'s access to y and $\sigma_4$'s second access to y are in conflict.

**Definition 4 (Schedule).** *A sequence of events $\sigma$ is called a* schedule *if all its events happen-before its last one, i.e., $e' \xrightarrow{\mathtt{hb}} e$ where $e$ is its last event, and $e'$ is any other event in $\sigma$. The last event $e$ of a schedule $\sigma$ is called the* head *of $\sigma$, sometimes denoted $hd(\sigma)$. For an execution sequence $E \cdot w$ and event $e \in w$, define the schedule $e \downarrow^w$ to be the subsequence $w'$ of $w$ such that (i) $e \in w'$, and (ii) for each $e' \in w$ it holds that $e' \in w'$ iff $e' \xrightarrow{\mathtt{hb}}_{E \cdot w} e$.*

### 4.1   Parsimonious Race Reversals

A central mechanism of many DPOR algorithms is to detect and reverse races. Intuitively, a race is a conflict between two consecutive accesses to a shared variable, where at least one access writes to the variable (i.e., it is a write or a read-modify-write).

**Definition 5 (Race).** *Let $E$ be an execution sequence. Two events $e$ and $e'$ in $E$ are* racing *in $E$ if (i) $e$ and $e'$ are performed by different threads, (ii) $e \xrightarrow{\mathtt{hb}}_E e'$. (iii) there is no other event $e''$ with $e \xrightarrow{\mathtt{hb}}_E e'' \xrightarrow{\mathtt{hb}}_E e'$.*

Intuitively, a race arises when two different threads perform dependent accesses to a shared variable, which are adjacent in the $\xrightarrow{\mathtt{hb}}_E$ order. If $e$ and $e'$ are racing in $E$, then to reverse the race, $E$ is decomposed as $E = E_1 \cdot e \cdot E_2$ with $e'$ in $E_2$, thereafter the schedule $\sigma = e' \downarrow^{E_2}$ is formed as the initial fragment of an alternative execution, which extends $E_1$.

The key idea of parsimonious race reversal is to reverse a race *only if* such a reversal generates an execution that has not been explored before. To be able to do so, POP remembers whenever an event in a new execution is in a schedule, and whether it is a schedule head. This can be done, e.g., by marking events in schedules, and specifically marking the schedule head. From now on, we consider such markings to be included in the events of executions. They play an important role in selecting races.

**Definition 6 (Fresh event).** *For an execution $E \cdot w \cdot e' \cdot w'$, the event $e'$ is called* fresh *in $w \cdot e' \cdot w'$ after $E$ if (i) if $e'$ is in a schedule, then it is the head of that schedule, and (ii) for each head $e_h$ of a schedule in $w$ it is the case that $e_h \xrightarrow{hb}_{E \cdot w \cdot e} e'$.*

**Definition 7 (Parsimonious race).** *Let $E$ be an execution sequence. Two events $e$ and $e'$ in $E$ are in a* parsimonious race, *denoted $e \lesssim_E e'$ if (i) $e$ and $e'$ are racing in $E$, (ii) $e$ is not in a schedule in $E$, and (iii) $e'$ is fresh in $w \cdot e'$ after $E_1$, where $E = E_1 \cdot e \cdot w \cdot e' \cdot w'$*

Conditions (ii) and (iii) are the additional conditions for a race to be parsimonious. They filter out races, whose reversals would lead to previously explored executions. Let us provide the intuition behind these conditions. (ii) If $e$ is in a schedule, then that schedule, call it $\sigma$, was generated by a race in an earlier explored execution $E'$. Hence $\sigma$ was contained in $E'$. Moreover $e'$ would race with the head of $\sigma$ also in $E'$; if $e'$ appeared after $\sigma$ the resulting new schedule had been generated already in $E'$; if $e'$ appeared before $\sigma$, then we would only undo a previous race reversal. This is illustrated in Fig. 2 by the race on y, between `a = y` and `y = 1` in $E_2$. (iii) If $e'$ is not fresh, then $e'$ appeared with the same happens-before predecessors in an earlier explored execution $E'$, where it was in a race that would generate the same schedule as in $E$. This is illustrated in Fig. 2 by the race on x, between `b = x` and `e = x`. in $E_2$, which was considered already in $E_1$.

### 4.2 The Parsimonious-OPtimal DPOR (POP) Algorithm

We will now describe the mechanism of the POP algorithm, without going into details regarding its handling of sleep sets (this will be done in Section 4.3). In particular, we will show how the *eager race reversal strategy* is represented in pseudo-code. Recall from Section 2 that a DPOR algorithm with parsimonious race reversal could be implemented so that the schedules that constructed from races with a particular event $e$ are all collected before they are explored. However, for some programs, the number of schedules created from races with an event $e$ can be exponential in the length of the longest program execution. In order not to consume exponential memory, POP explores schedules *eagerly*: immediately after the creation of a schedule, exploration switches to continuations of that schedule.

The POP algorithm is shown as Algorithm 1, where lines without background shading are concerned with the exploration and race handling, and the other lines, which are marked with green background, are concerned with sleep sets. POP takes an input program, and explores its executions by repeated calls to the procedure `Explore`. For each prefix $E'$ of an execution that is under exploration, the algorithm maintains a characterization $SSChar[E']$ of the sleep set at $E'$, to be described in Section 4.3, in order to prevent redundant exploration of read schedules. This characterization is manipulated by POP through two functions:

`MkSchedChar`$(\sigma, E_1, e, E_2)$ constructs a characterization of the sleep set for a newly constructed $\sigma$, constructed from a race $e \lesssim_{E_1 \cdot e \cdot E_2} \texttt{last}\,(E_2)$,

**Algorithm 1:** POP (Recursive)

1   $SSChar[\langle\rangle] = \emptyset$
2   `Explore`$(\langle\rangle)$
3   `Explore`$(E)$
4     **foreach** $e$ s.t. $e \lesssim_E e'$, where $e' = \texttt{last}(E)$ **do**
5       **let** $E = E_1 \cdot e \cdot E_2$
6       **let** $\sigma = e' \downarrow^{E_2}$
7       **if** $\texttt{UpdSeq}(\sigma, SSChar[E_1]) \neq block$ **then**
8         $SSChar[E_1 \cdot \sigma] = \texttt{UpdSeq}(\sigma, SSChar[E_1])$
9         **if** $(e'.\texttt{T} = \texttt{R})$ **then**
10           $SchedChar[E_1](\sigma) = \texttt{MkSchedChar}(\sigma, E_1, e, E_2)$
11           $SSChar[E_1 \cdot \sigma] \cup= SchedChar[E_1](\sigma)$
12         `Explore`$(E_1 \cdot \sigma)$
13     **if** $\exists e \in \texttt{enabled}(E)$ s.t. $\texttt{UpdSeq}(\langle e\rangle, SSChar[E]) \neq block$ **then**
14       $SSChar[E \cdot e] = \texttt{UpdSeq}(\langle e\rangle, SSChar[E])$
15       `Explore`$(E \cdot e)$

$\texttt{UpdSeq}(w, SSChar)$ updates the sleep set characterization $SSChar$ wrt. processing of the sequence $w$. However, if a characterized read schedule (i.e., a schedule in the sleep set) would be performed while exploring $w$, the function returns *block* instead of the updated characterization.

The algorithm initializes the characterizations of sleep sets of $\langle\rangle$ (line 1) and calls `Explore`$(\langle e\rangle)$ (line 2). Each call to `Explore`$(E)$ consists of a race reversal phase (lines 4 to 12) and an exploration phase (lines 13 to 15). In the race reversal phase, POP considers all parsimonious races between an event $e$ in $E$ and the last event $e'$ of $E$ (line 4). For each such race, of form $e \lesssim_E e'$, POP decomposes $E$ as $E_1 \cdot e \cdot E_2$ (line 5), and forms the schedule $\sigma$ that reverses the race as $e' \downarrow^{E_2}$ (line 6). It then intends to call `Explore`$(E_1 \cdot \sigma)$ in order to recursively switch the exploration to the newly reversed race, according to the eager race reversal strategy. Before that it checks whether exploring $E_1 \cdot \sigma$ will complete a schedule in the sleep set by calling $\texttt{UpdSeq}(\sigma, SSChar[E_1])$ (line 7). If not, $SSChar[E_1 \cdot \sigma]$ is computed (line 8), and if $e'$ is a read event also extended with the new sleep set for $\sigma$ (lines 10 to 11). After these preparations, `Explore`$(E_1 \cdot \sigma)$ is called recursively (line 12). After the return of all recursive calls initiated in the race reversal phase, `Explore` enters the exploration phase. There it picks an event $e$ that is enabled for execution, and check that $e$ is not the head of a schedule in the sleep set by calling $\texttt{UpdSeq}(\langle e\rangle, SSChar[E])$ (line 13) If the check succeeds, exploration of $e$ is prepared by updating $SSChar[E_1 \cdot e]$ (line 14) and then performed by calling `Explore`$(E \cdot e)$(line 15).

We establish (in Lemma 1) that the recursion depth of Algorithm 1 is at most $n(n-1)/2$, where $n$ is the length of the longest execution of the analyzed program.

### 4.3   Parsimonious Sleep Set Characterization

As described in Section 2, POP needs a sleep set mechanism to avoid redundant exploration of read schedules. Such a mechanism is needed whenever POP explores reversals of races with a write event $e_W$ that appears after an execution $E_1$. Then each parsimonious race $e_W \lesssim_{E'} e_R$ between $e_W$ and a read event $e_R$ results in a schedule $\sigma$, which will be explored as a continuation of $E_1$. For any two such schedules, $\sigma$ and $\sigma'$, POP must ensure that *either* the exploration starting with $\sigma$ does not continue in a way that includes $\sigma'$, *or* (vice versa) that the exploration starting with $\sigma'$ does not continue in a way that includes $\sigma$. In Section 2, it was further described that to achieve this, POP must for each such explored write event $e_W$ establish a total order between the read schedules resulting from races with $e_W$, and ensure that an exploration starting with $\sigma$ does not continue in a way that includes another schedule $\sigma'$ which precedes $\sigma$ in this order. It was also observed that, since there can be an exponential number of such schedules, the naïve approach of enumerating the schedules that precede $\sigma$ can in the worst case consume space exponential in the length of the longest execution.

In this section, we will describe one way to realize such a sleep set mechanism. We first define, for each explored write event $e_W$, a total order between the read-schedules resulting from races with $e_W$. Thereafter we define a succinct (polynomial-space) characterization of all schedules that precede any given such schedule $\sigma$. Finally, we define a polynomial-space mechanism for POP to monitor exploration so that exploration after the schedule $\sigma$ does not explore another read schedule which precedes $\sigma$ in the order.

First, for a variable $x$, we define a *read-x-schedule* to be a schedule whose head is a read on $x$, and which does not contain any other read or write on $x$. A *read-schedule* is a *read-x-schedule* for some variable $x$. Then a read-$x$-schedule is a schedule that may be formed when reversing a parsimonious race between a write on $x$ and a read on $x$. Such a schedule $\sigma$ cannot include a write on $x$, since then it could not have been formed from a race. Also, it cannot include a read on $x$, since that extra read will both happen-before $hd(\sigma)$, and happen-after the write on $x$, contradicting that there was a race between the write and $hd(\sigma)$.

Let us now define the order $\propto$, which for each write event $e_W$ totally orders the schedules that result from parsimonious races between $e_W$ and a subsequent read. Let $\sigma$ be formed from a race $e_W \lesssim_E e_R$ between $e_W$ and another read event $e_R$ in $E$ and $\sigma'$ be formed from a race $e_W \lesssim_{E'} e'_R$ between $e_W$ and another read event $e'_R$. Then $\sigma' \propto \sigma$ if either

(A) $E'$ is a prefix of $E$, i.e., $e'_R$ occurs before $e_R$ in $E$, or
(B) for the longest common prefix $\hat{E}$ of $E$ and $E'$, $E$ has a prefix of form $\hat{E} \cdot \hat{e}$ for some non-schedule event $\hat{e}$, whereas $E'$ has a prefix of form $\hat{E} \cdot \hat{\sigma}$ for some schedule $\hat{\sigma}$ (which is induced by a race whose first event is $\hat{e}$), or
(C) for the longest common prefix $\hat{E}$ of $E$ and $E'$, $E$ has a prefix of form $\hat{E} \cdot \sigma_i$ for some schedule $\sigma_i$, whereas $E'$ has a prefix of form $\hat{E} \cdot \sigma'_i$ for some schedule $\sigma'_i$, and $\sigma'_i \propto \sigma_i$.

Schedules of form (A) are called *contained schedules* (wrt. $\sigma$). An example can be found in Fig. 2. Consider the schedules $\sigma_3 := \langle c = \mathtt{y} \cdot \mathtt{y} = 1 \cdot \mathtt{z} = 1 \cdot d = \mathtt{z} \cdot e = \mathtt{x} \rangle$ from the race $\mathtt{x=1} \lesssim_{E_3} e=\mathtt{x}$ in $E_3$, and $\sigma_4 := \langle c=\mathtt{y} \cdot \mathtt{y}=1 \cdot \mathtt{g}=1 \cdot a=\mathtt{y} \cdot b=\mathtt{x} \rangle$ from the race $\mathtt{x=1} \lesssim_{E_3} b=\mathtt{x}$ in $E_3$. As $e=\mathtt{x}$ occurs before $b=\mathtt{x}$ in $E_3$, (A) implies that $\sigma_3 \propto \sigma_4$. Schedules of form (B) are called *conflicting schedules*, beacuse $e'$ occurs in an execution which branches off from (thus conflicts with) $E$ because of a race involving an event $\hat{e}$ in $E$. For example, consider the schedules $\sigma_2 := \langle c=\mathtt{y} \cdot d=\mathtt{z} \cdot e=\mathtt{x} \rangle$, which is constructed from the race between $\mathtt{x=1}$ and $e=\mathtt{x}$ in $T_5$, and $\sigma_4$ constructed from the race $\mathtt{x=1} \lesssim_{E_3} b=\mathtt{x}$ in $E_3$. Since $T_5$ branches off from (and thus conflicts with) $E_3$ after the prefix $\hat{E} := \langle \mathtt{x=1} \cdot c=\mathtt{y} \cdot \mathtt{y}=1 \rangle$ with the schedule $\hat{\sigma} := \langle d=\mathtt{z} \rangle$, we have $\sigma_2 \propto \sigma_4$ according to case (B). Schedules of form (C) are called *inherited schedules*, because the order $\sigma' \propto \sigma$ is inherited from the order $\sigma'_i \propto \sigma_i$. For example, consider the schedules $\langle \mathtt{g}=1 \cdot a=\mathtt{y} \rangle$ (second branch after $\mathtt{x=1}$), and $\langle c=\mathtt{y} \rangle$ (third branch after $\mathtt{x=1}$), for which $\langle \mathtt{g}=1 \cdot a=\mathtt{y} \rangle \propto \langle c=\mathtt{y} \rangle$ because of (A). Now consider the schedules $\sigma_1 := \langle \mathtt{g}=1 \cdot a=\mathtt{y} \cdot b=\mathtt{x} \rangle$ from the race $\mathtt{x=1} \lesssim_{E_2} b=\mathtt{x}$ in $E_2$, and $\sigma_2$ from the race between the events $\mathtt{x=1}$ and $e=\mathtt{x}$ from an execution in $T_5$. As $\langle \mathtt{g}=1 \cdot a=\mathtt{y} \rangle$ is a prefix of $\sigma_1$ and $\langle c=\mathtt{y} \rangle$ is a prefix of $\sigma_2$, according to (C), the order $\langle \mathtt{g}=1 \cdot a=\mathtt{y} \rangle \propto \langle c=\mathtt{y} \rangle$ is inherited as $\sigma_1 \propto \sigma_2$.

It is clear that these rules define a total order on the read schedules that branch off after $E_1$. We next define a succinct way to characterize, for a given schedule $\sigma$, the set of schedules $\sigma'$ such that $\sigma' \propto \sigma$. Given $E = E_1 \cdot e_W \cdot w \cdot e_R$ and $\sigma$ formed from the race $e_W \lesssim_E e_R$, let $w = w_0 \cdot \sigma_1 \cdot w_1 \cdot \sigma_2 \cdots \sigma_m \cdot w_m$, where $\sigma_1, \ldots, \sigma_m$ are the schedules in $w$. We note that $\sigma$, since $e_W \lesssim_E e_R$ is parsimonious, includes all $\sigma_i$ (including their heads) for $1 \leq i \leq m$, and may also include events in the sequences $w_0, \ldots, w_m$. This means that $w \backslash \sigma$ is of form $w'_0 \cdot \ldots \cdot w'_m$, where $w'_i$ is the sequence remaining in $w_i$ after removing $\sigma$; in particular $w \backslash \sigma$ does not contain any events in any schedule $\sigma_i$. The following proposition characterizes how to detect a schedule $\sigma'$ with $\sigma' \propto \sigma$ in an exploration that is initiated as a continuation of $E_1 \cdot \sigma$.

**Proposition 1.** *Let $E = E_1 \cdot e_W \cdot w \cdot e_R$, let $w = w_0 \cdot \sigma_1 \cdot w_1 \cdots w_m$, and let $\sigma$ be formed from $e_W \lesssim_E e_R$. Let $w'_j = w_j \backslash \sigma$ for $j = 0, \ldots, m$, and $e_j = hd(\sigma_j)$ for $j = 1, \ldots, m$. Let $E_1 \cdot \sigma \cdot u \cdot e'_R$ be an execution where $e'_R$ is a read event on $x$, and let $\sigma' = e'_R \downarrow^{\sigma \cdot u \cdot e'_R}$. Then $\sigma' \propto \sigma$ iff $\sigma'$ is a read-$x$-schedule such that either*

(A) (i) $(e'_R \downarrow^{u \cdot e'_R}) \sqsubseteq w \backslash \sigma$, and (ii) if $e'_R$ is in $w'_j$ then $e_k \xrightarrow{\text{hb}} e'_R$ for $1 \leq k \leq j$,
(B) for some $j$ with $0 \leq j \leq m$ we have (i) $(e'_R \downarrow^{u \cdot e'_R}) \not\propto w'_0 \cdot \ldots \cdot w'_j$, and
   (ii) if $j$ is the smallest index s.t. (i) holds, then $e_k \xrightarrow{\text{hb}} e'_R$ for $1 \leq k \leq j$, or
(C) for some $i$ with $1 \leq i \leq m$ s.t. $\sigma_i$ is a read-schedule, and $\sigma'_i$ with $\sigma'_i \propto \sigma_i$
   1) if $hd(\sigma_i).var \neq x$ then (i) $(hd(\sigma'_i) \downarrow^{w'_0 \cdots w'_i \cdot \sigma'_i}) \sqsubseteq u$,
      (ii) $hd(\sigma'_i) \xrightarrow{\text{hb}} e'_R$, and (iii) $e_k \xrightarrow{\text{hb}} e'_R$ for $1 \leq k \leq i$.
   2) if $hd(\sigma_i).var = x$ then (i) $(hd(\sigma'_i) \downarrow^{w'_0 \cdots w'_i \cdot \sigma'_i}) \sqsubseteq u \cdot e'_R$,
      (ii) $hd(\sigma'_i) = e'_R$, and (iii) $e_k \xrightarrow{\text{hb}} e'_R$ for $1 \leq k \leq i$.  □

Let us motivate this proposition.

(A) Since $\sigma \sqsubseteq w \cdot e_R$, condition (i) implies that $\sigma' = e'_R \downarrow^{\sigma \cdot u \cdot e'_R} \sqsubseteq w$, implying that $\sigma'$ is a *contained schedule* (wrt. $\sigma$).

(B) Since $\sigma \sqsubseteq w \cdot e_R$, condition (i) implies that $\sigma' = e'_R \downarrow^{\sigma \cdot u \cdot e'_R} \not\propto w_0 \cdot \sigma_1 \cdot \ldots \cdot \sigma_j \cdot w_j$, implying that $\sigma'$ is a *conflicting schedule*.

(C) Let us first consider case 1). Since $\sigma \sqsubseteq w \cdot e_R$, condition (i) implies that $(hd(\sigma'_i) \downarrow^{w_0 \cdot \sigma_1 \cdot \ldots \cdot \sigma_j \cdot w_j \cdot \sigma'_i}) \sqsubseteq \sigma \cdot u$, implying that $\sigma'$ is an *inherited schedule*. Condition (ii) ensures that $e'_R$ appears in the exploration that follows $hd(\sigma'_i)$, in which case $hd(\sigma'_i) \xrightarrow{\text{hb}} e'_R$ is necessary for $e'_R$ to be fresh. Case 2) is a slight modification for the case that the head of $\sigma'_i$ and $e'_R$ read from the same variable, in which case $e'_R$ must be $hd(\sigma'_i)$ (since a read-$x$-schedule cannot contain another read on $x$).

In each case, the last condition ensures that $e'_R$ is fresh, and thus part of a parsimonious race.

Let us illustrate, using Fig. 2, how some continuations of read schedules can be characterized according to Proposition 1. First, consider $\sigma_4$ (top right in Fig. 2), derived from the race $\mathtt{x\,=\,1} \lesssim_{E_3} b\,\mathtt{=\,x}$ in $E_3$. Decomposing $E_3$ as $\mathtt{x\,=\,1} \cdot w \cdot b\,\mathtt{=\,x}$, where $w := \langle c\,\mathtt{=\,y} \cdot \mathtt{y\,=\,1} \cdot \mathtt{z\,=\,1} \cdot d\,\mathtt{=\,z} \cdot e\,\mathtt{=\,x} \cdot \mathtt{g\,=\,1} \cdot a\,\mathtt{=\,y} \rangle$, we obtain $w \setminus \sigma_4 = \langle \mathtt{z\,=\,1} \cdot d\,\mathtt{=\,z} \cdot e\,\mathtt{=\,x} \rangle$.

(A) Assume that the exploration continues after $\sigma_4$ as $\langle \mathtt{z\,=\,1} \cdot d\,\mathtt{=\,z} \cdot e\,\mathtt{=\,x} \rangle$. Letting $u$ be $\langle \mathtt{z\,=\,1} \cdot d\,\mathtt{=\,z} \rangle$ and $e'_R$ be $e\,\mathtt{=\,x}$, we see that $\sigma' = u \cdot e'_R$ matches the conditions in case (A), since (i) $\sigma' \sqsubseteq w \setminus \sigma_4$ and (ii) $e'_R$ happens-after the head of the only schedule $\langle c\,\mathtt{=\,y} \rangle$ in $E_3$.

(B) Assume next that the exploration continues after $\sigma_4$ as $\langle d\,\mathtt{=\,z} \cdot e\,\mathtt{=\,x} \rangle$. Letting $u$ be $\langle d\,\mathtt{=\,z} \rangle$ and $e'_R$ be $e\,\mathtt{=\,x}$, we see that $\sigma' = u \cdot e'_R$ matches the conditions in case (B), since (i) $e\,\mathtt{=\,x} \downarrow^{\langle d\,\mathtt{=\,z}\cdot e\,\mathtt{=\,x}\rangle} = \langle d\,\mathtt{=\,z} \cdot e\,\mathtt{=\,x} \rangle$ and $\langle d\,\mathtt{=\,z} \cdot e\,\mathtt{=\,x} \rangle \not\propto (w \setminus \sigma_4)$, and (ii) $e\,\mathtt{=\,x}$ happens-after the head of the only schedule $c\,\mathtt{=\,y}$ in $E_3$.

(C) Let us next consider $\sigma_2$ (top middle in Fig. 2), derived from the race $\mathtt{x\,=\,1} \lesssim_{E'} e\,\mathtt{=\,x}$ in the first explored execution $E'$ from $T_5$. Decomposing $E'$ as $\mathtt{x\,=\,1} \cdot w \cdot e\,\mathtt{=\,x}$, where $w := \langle c\,\mathtt{=\,y} \cdot \mathtt{y\,=\,1} \cdot d\,\mathtt{=\,z} \rangle$ we obtain $w \setminus \sigma_2 = \langle \mathtt{y\,=\,1} \rangle$. Assume next that the exploration continues after $\sigma_2$ as $\langle \mathtt{g\,=\,1} \cdot a\,\mathtt{=\,y} \cdot b\,\mathtt{=\,x} \rangle$. Letting $u$ be $\langle \mathtt{g\,=\,1} \cdot a\,\mathtt{=\,y} \rangle$ and $e'_R$ be $b\,\mathtt{=\,x}$, we see that $\sigma' = u \cdot e'_R$ matches the conditions in case (C)1), since there is the schedule $\sigma_i = \langle c\,\mathtt{=\,y} \rangle$ for which there is another schedule $\sigma'_i = \langle \mathtt{g\,=\,1} \cdot a\,\mathtt{=\,y} \rangle$ with $\sigma'_i \propto \sigma_i$. The conditions in case (C)1) are satisfied, since (i) $a\,\mathtt{=\,y} \downarrow^{\langle \mathtt{y\,=\,1}\cdot \mathtt{g\,=\,1}\cdot a\,\mathtt{=\,y}\rangle} \sqsubseteq u$, (ii) $a\,\mathtt{=\,y} \xrightarrow{\text{hb}} e'_R$, and (iii) there is no schedule before the event $c\,\mathtt{=\,y}$ in $E'$.

Based on Proposition 1, we now describe a technique to monitor the exploration of executions in order to detect when it is about to explore a schedule in a sleep set. It is based on annotating each newly constructed read schedule $\sigma$ with a characterization of the schedules $\sigma'$ with $\sigma' \propto \sigma$ that must be avoided in the exploration that continues after $\sigma$. We use the same notation and set-up as for Proposition 1. For $i = 0, \ldots, m$, let $P_i$ denote $P_i = w'_0 \{e_1\} w'_1 \cdots \{e_i\} w'_i$, where $w'_j$ is $w_j \setminus \sigma$ for $j = 0, \ldots, i$, and $e_j$ is $hd(\sigma_j)$ for $j = 1, \ldots, i$. From Proposition 1 we see that (i) $P_m$ and $x$ contains sufficient information to characterize the contained and conflicting schedules that must be avoided, and (ii) for each $i = 1, \ldots, m$, such that $\sigma_i$ is a read-schedule, $P_{i-1}$ together with a characterization of the schedules $\sigma'_i$ with $\sigma'_i \propto \sigma_i$ contain sufficient information to characterize the schedules inherited from schedules $\sigma'_i$ with $\sigma'_i \propto \sigma_i$ that must be avoided. Let us therefore define a *schedule expression* as an expression of form (i) $P_m \rhd x$, characterizing the set of contained and conflicting read-$x$-schedules, according to cases (A) and (B) in Proposition 1, or of form (ii) $P_{i-1}[\varphi_i] \rhd x$ for some $i = 1, \ldots, m$, such that $\sigma_i$ is a read-schedule, and $\varphi_i$ is a schedule expression characterizing schedules $\sigma'_i$ with $\sigma'_i \propto \sigma_i$. Let us go through one example of each form of schedule expressions using Fig. 2. While exploring continuations of $\sigma_4$, POP creates two schedule expressions; (i) $P_1 \rhd x = \{c\texttt{=}y\}\langle z\texttt{=}1 \cdot d\texttt{=}z \cdot e\texttt{=}x\rangle \rhd \texttt{x}$ representing the schedules $\sigma_2$ and $\sigma_3$, and (ii) $P_0[\varphi_1] \rhd x = \langle\rangle[\langle z\texttt{=}1 \cdot g\texttt{=}1 \cdot a\texttt{=}y \cdot b\texttt{=}x\rangle \rhd \texttt{y}] \rhd \texttt{x}$, representing only $\sigma_1$. Notice that, expression (ii) is useless in this case as $\sigma_1$ is conflicting with $\sigma_4$, i.e., $\sigma_1$ is not a feasible continuation after $\sigma_4$. However, the same expression is useful to prevent doing $\sigma_1$, when exploring a continuation of $\sigma_2$.

In order to detect when exploration is about to explore a schedule that must be avoided, the "state" of each schedule expression will during exploration be maintained by POP in a *sleep set expression*, which is obtained from a schedule expression $\varphi$ by (i) augmenting each event $e$ which occurs in some sequence $w_i$ in $\varphi$ (i.e., not inside brackets $\{\cdot\}$) with a *conflict set* (denoted $C$) of encountered events that conflict with $e$ or happen-after an event that conflicts with $e$; we use the notation $e^C$ to denote such an augmented event, (ii) augmenting each enclosed subexpression of form $P \rhd x$ or $P[\varphi] \rhd x$ with the set (denoted $D$) of encountered read-$x$-events that are heads of read-schedules that are characterized by $P \rhd x$; we use the notation $[P \rhd x]^D$ (or $[P[\varphi] \rhd x]^D$), and (iii) augmenting each occuring variable $x$ that occurs after $\rhd$ in a subexpression of form $P \rhd x$ or $P[\varphi] \rhd x$ with the set of previously encountered read events on $x$; we use the notation $\rhd x^R$, where $R$ is this set of read events. If a read on $x$ happens-after a read in $R$, it cannot be the head of a read-$x$-schedule, and should thus not be blocked (recall from the definition of read-$x$-schedules that its head cannot happen-after another read on the same variable). When a sleep set expression is created and initialized, its augmenting sets are empty. We identify a schedule expression with its initialized sleep set expression. We use $\psi$, possibly with sub- or superscripts, to range over sleep set expressions.

Algorithm 2 shows POP's implementation of the sleep set expression manipulation functions $\texttt{MkSchedChar}(\sigma, E_1, e, E_2)$ and $\texttt{UpdSeq}(w, SSChar)$, which are called by Algorithm 1. A set of sleep set expressions is called a *sleep set characterization*. The function $\texttt{MkSchedChar}(\sigma, E_1, e, E_2)$ (line 1), constructs the set of schedule expressions (which can be seen as initialized sleep set expressions) for $\sigma$ according to the description given earlier in this section. The function $\texttt{UpdSeq}(w, SSChar)$ updates the sleep set characterization $SSChar$ wrt. processing of the sequence $w$. At its top level, $\texttt{UpdSeq}(w, SSChar)$ updates each sleep set expression $\psi$ in $SSChar$ with the sequence of events in $w$, one by one, each time calling $\texttt{UpdSE}(e, \psi)$, where $\psi$ is of the form $P \rhd x^R$ or $\psi = P[\psi']^D \rhd x^R$. In the pseudo code, we used $\psi.var$ to denote $x$, the variable of the expression. The function $\texttt{UpdSE}(e, \psi)$ returns a pair $\langle \psi', res\rangle$, where $\psi'$ is the updated expression denoted by $\texttt{UpdSE}(e, \psi).exp$ and $res$ is the result denoted by $\texttt{UpdSE}(e, \psi).res$. The result $\texttt{UpdSE}(e, \psi).res$ is *block* when $\psi$ characterizes a read-$x$-schedule in the sleep set whose head is $e$, *indep* when $e$ does not update $\psi$, and *continue* otherwise. If $e$

**Algorithm 2:** Functions for Parsimonious Sleep Set Characterization

1    MkSchedChar($\sigma, E_1, e, E_2$)
2      **let** $x = hd(\sigma).var$
3      **let** $E_2 = w_0 \cdot \sigma_1 \cdot w_1 \cdot \cdots \cdot \sigma_m \cdot w_m \cdot e'$
4      **for** $i = 0, \ldots, m$ **let** $P_i = (w_0 \backslash \sigma)\{hd(\sigma_1)\}(w_1 \backslash \sigma) \cdots \{hd(\sigma_i)\}(w_i \backslash \sigma)$
5      **for** $i = 0, \ldots, m$ **let** $u_i = e \cdot w_0 \cdot \sigma_1 \cdot \ldots \cdot \sigma_i \cdot w_i$
6      **return** $\left( \{P_m \triangleright x\} \cup \bigcup_{i=1}^{m} \{P_{i-1}[\varphi_i] \triangleright x \mid \varphi_i \in SchedChar[E_1 \cdot u_{i-1}](\sigma_i)\} \right)$

7    UpdSeq($w, SSChar$)
8      **if** $w = \langle \rangle$ **then return** $(SSChar)$
9      **else if** $\exists \psi \in SSChar : \text{UpdSE}(fst(w), \psi).res = block$ **then return** $block$
10     **else return** UpdSeq($rest(w), \{\text{UpdSE}(fst(w), \psi).exp \mid \psi \in SSChar \wedge$
11                    **if** $fst(w).\text{T} = \text{W}$ **then** $\psi$ does not contain $\triangleright(fst(w).var)\})$

12    UpdSE($e, P \triangleright x^R$)
13      **return** UpdP($e, P, x, R$)

14    UpdSE($e, P[\psi']^D \triangleright x^R$)
15      **if** UpdP($e, P, x, R$).$res = indep$ **then**
16        **let** $P = w'_0\{e'_1\}w'_1 \cdots \{e'_i\}w'_i$
17        **if** $(e.\text{T} = \text{R} \wedge e.var = x \wedge (\exists e' \in D : [e'.var \neq x \wedge e' \xrightarrow{hb} e])$ **then**
18          **if** $\nexists e'_R \in R : e'_R \xrightarrow{hb} e \wedge \forall n : 1 \leq n \leq i : e'_n \xrightarrow{hb} e$ **then**
19            **return** $\left\langle P[\psi']^D \triangleright x^{R \cup \{e\}}, block \right\rangle$
20          **else return** $\left\langle P[\psi']^D \triangleright x^{R \cup \{e\}}, continue \right\rangle$
21        **else if** UpdSE($e, \psi'$).$res = block$ **then**
22          **if** $x = \psi'.var$ **then**
23            **if** $\nexists e'_R \in R : e'_R \xrightarrow{hb} e \wedge \forall n : 1 \leq n \leq i : e'_n \xrightarrow{hb} e$ **then** $res = block$
24            **else** $res = continue$
25            **return** $\left\langle P[\text{UpdSE}(e, \psi').exp]^{D \cup \{e\}} \triangleright x^{R \cup \{e\}}, res \right\rangle$
26          **else return** $\left\langle P[\text{UpdSE}(e, \psi').exp]^{D \cup \{e\}} \triangleright x^R, continue \right\rangle$
27        **else return** $\left\langle P[\text{UpdSE}(e, \psi').exp]^D \triangleright x^R, \text{UpdSE}(e, \psi').res \right\rangle$
28      **else return** $\left\langle \text{UpdP}(e, P, x, R).exp[\psi']^D \triangleright x^R, \text{UpdP}(e, P, x, R).res \right\rangle$

29    UpdP($e, P, x, R$)
30      **let** $P = w'_0\{e'_1\}w'_1 \cdots \{e'_i\}w'_i$
31      **for** $j = 0, \ldots, i$ **do**
32        **let** $w'_j = e_1^{C_1} \cdot \ldots \cdot e_k^{C_k}$
33        **for** $l = 1, \ldots, k$ **do**
34          **if** $e \bowtie C_l \vee e \leftrightarrow e_l$ **then** add $e$ to $C_l$ ; **go to** line 37
35          **if** $e = e_l$ **then** remove $e$ from $w'_j$ ; **go to** line 37
36      **return** $\langle P, indep \rangle$
37      **if** $(e.\text{T} = \text{R} \wedge e.var = x \wedge \forall n : 1 \leq n \leq j : e'_n \xrightarrow{hb} e)$ **then**
38        **if** $\nexists e'_R \in R : e'_R \xrightarrow{hb} e$ **then return** $\langle$the updated version of $P, block\rangle$
39        **else** add $e$ to $R$
40      **else return** $\langle$the updated version of $P, continue\rangle$

is a write on a variable $y$, then in this process, all sleep set expressions containing $\triangleright y$ are discarded, since it is from now impossible to complete a read-$y$-schedule in the sleep set characterized by $\psi$. The function $\texttt{UpdSE}(e, \psi)$ comes in two versions (at line 12 and line 14). Both versions first call $\texttt{UpdP}(e, P, x, R)$, which updates the sleep set expressions with respect to contained and conflicting read-$x$-schedules characterized by $P \triangleright x^R$. Like the $\texttt{UpdSE}$ function, $\texttt{UpdP}(e, P, x, R)$ returns the updated version of $P$ and the result *block* if $e$ is the head of such a schedule, *indep* if $e$ is independent with all of $P$, *continue* otherwise. In the code for $\texttt{UpdP}$, we let let $e \leftrightarrow e'$ denote that $e$ and $e'$ are performed by *different* threads accessing the same variable and at least one of $e$ and $e'$ writes. For an event $e$ and a set $C$ of events, let $e \bowtie C$ denote that there is some $e' \in C$ with $e \bowtie e'$. When called, $\texttt{UpdP}(e, P, x, R)$ traverses the sequences $w'_0, \ldots, w'_i$, one event at a time, and stops at the first event $e_l^{C_l}$ such that either (i) $e \leftrightarrow e_l$ or $e \bowtie C_l$, in which case $e$ is added to $C_l$ (line 34), or (ii) $e = e_l$, in which case $e$ is removed from the sequence (of form $w'_j$) (line 35). If in addition $e$ is a read on $x$ and happens after the relevant schedule heads among $e'_1, \ldots, e'_i$ for being fresh, then if $e$ does not happen after a read in $R$, $\texttt{UpdP}(e, P, x, R)$ returns *blocked* (since in case (i) it is the head of a conflicting schedule and in case (ii) of a contained schedule), else $e$ is added to $R$ (line 39). If on the other hand, $e$ is not a read on $x$ or does not happen after the relevant schedule heads among $e'_1, \ldots, e'_i$, then $\texttt{UpdP}(e, P, x, R)$ returns the updated version of $P$ and *continue*. Finally, if there is no event $e_l^{C_l}$ in $w'_0, \ldots, w'_i$ satisfying conditions (i) or (ii), then $\texttt{UpdP}(e, P, x, R)$ returns *indep* (line 36).

Let us now consider $\texttt{UpdSE}(e, \psi)$, which comes in two versions, depending on the form of $\psi$. If $\psi$ is of form $P \triangleright x^R$ (line 12), it calls $\texttt{UpdP}(e, P, x, R)$ and forwards its return value. If $\psi$ is of form $P[\psi']^D \triangleright x^R$ (line 14), it also calls $\texttt{UpdP}(e, P, x, R)$. Also, this version forwards the return value when $\texttt{UpdP}(e, P, x, R)$ returns *block* or *continue*. In addition, if $\texttt{UpdP}(e, P, x, R)$ returns *indep*, meaning that $e$ is independent of $P$, then

(i) if some event already in $D$ (being the head of a schedule characterized by $\psi'$) happens-before $e$ and $e$ is a read on $x$, then $e$ is added to $R$ and the updated $P$ is returned with either the result *block* when $e$ does not happen after a read in $R$ and $e$ happens after all the schedule heads in $P$ (at line 19 since $e$ is the head of an inherited schedule), or the result *continue* otherwise (line 20),
(ii) else if $\texttt{UpdSE}(e, \psi')$ returns *block*, then if $x$ is same as the variable of $\psi'$ (line 22), $e$ is added to the sets $R$ and $D$, and (a) either the function returns *block* (line 23, line 25) when $\psi'$ characterizes a read-$\psi'.var$-schedule (Case (C) (2) of Proposition 1), (b) or the function returns *continue* (line 24, line 25), else $e$ is added to $D$ and the function returns *continue* (line 26),
(iii) else the inner sleep set expression $\psi'$ is updated by calling $\texttt{UpdSE}(e, \psi')$ and returned (line 27).

Finally, if $\texttt{UpdP}(e, P, x, R)$ returns neither *blocked* nor *indep*, the updated sleep set expression is returned (line 28).

## 5  Correctness and Space Complexity

In this section, we state theorems of correctness, optimality, and space complexity of POP. We first consider correctness and optimality.

**Theorem 1.** *For a terminating program P, the POP algorithm has the properties that (i) for each maximal execution E of P, it explores some execution $E'$ with $E' \simeq E$, (ii) it never explores two different but equivalent maximal executions, and (iii) it is never blocked (at line 13) unless it has explored a maximal execution.*

We thereafter consider space complexity.

**Lemma 1.** *The number of nested recursive calls to* Explore *at line 12 is at most $n(n-1)/2$, where n is the length of the longest execution of the program.*

Note that in this lemma, we do not count the calls at line 15, since they are considered as normal exploration of some execution. Only the calls at line 12 start the exploration of a new execution.

**Theorem 2.** *Algorithm 1 needs space which is polynomial in n, where n is the length of the longest execution of the analyzed program.*

## 6  Implementation and Evaluation

Our implementation, which is available in the artifact of this paper, was done in a fork of NIDHUGG. NIDHUGG is a state-of-the-art stateless model checker for C/C++ programs with Pthreads, which works at the level of LLVM Intermediate Representation, typically produced by the Clang compiler. NIDHUGG comes with a selection of DPOR algorithms, one of which is Optimal DPOR [1] nowadays also enhanced with Partial Loop Purity elimination and support for await statements [25]. In our NIDHUGG fork, we have added the POP algorithm as another selection. Its implementation involved: (i) designing an efficient data structure to simulate recursive calls to Explore, i.e., follow the next schedule to explore and backtrack to the previous execution when no further races to reverse, (ii) developing a procedure to filter out races that are not parsimonious, and (iii) implementing a more optimized data structure than Algorithm 2 that stores sleep set characterizations as trees.

In this section, we evaluate the performance of POP's implementation and compare it, in terms of time and memory, against the implementations of Optimal DPOR in NIDHUGG commit 5805d77 and the graph-based Truly Stateless (TruSt) Optimal DPOR algorithm [29] as implemented in GENMC v0.10.0 using options -sc --disable-instruction-caching. All tools employed LLVM 14.0.6, and the numbers we present are measured on a desktop with a Ryzen 7950X CPU running Debian 12.4.

Table 1 contains the results of our evaluation. Its first nine benchmarks are from the DPOR literature, and are all parametric on the number of threads (shown in parentheses). The last benchmark, length-param, is synthetic and is additionally parametric on the length of its executions. Since these DPOR algorithms are optimal, they explore the same number of executions (2nd column) in all ten benchmarks. We will analyze the results in five groups (cf. Table 1).

**Table 1.** Time and memory performance of three optimal DPOR algorithms on ten benchmark programs which are parametric in the number of threads used.

| Benchmark | Executions | Time (secs) | | | Memory (MB) | | |
|---|---|---|---|---|---|---|---|
| | | TruSt | Optimal | POP | TruSt | Optimal | POP |
| circular-buffer(7) | 3432 | 0.62 | 0.45 | 0.43 | 85 | 84 | 84 |
| circular-buffer(8) | 12870 | 2.63 | 1.79 | 1.66 | 85 | 84 | 84 |
| circular-buffer(9) | 48620 | 11.04 | 7.21 | 6.67 | 85 | 84 | 84 |
| fib-bench(4) | 19605 | 1.08 | 1.93 | 1.82 | 85 | 84 | 84 |
| fib-bench(5) | 218243 | 14.59 | 24.66 | 24.10 | 85 | 84 | 84 |
| fib-bench(6) | 2364418 | 186.25 | 301.30 | 297.40 | 85 | 84 | 84 |
| linuxrwlocks(6) | 99442 | 3.61 | 13.71 | 12.88 | 90 | 91 | 91 |
| linuxrwlocks(7) | 829168 | 32.75 | 127.66 | 121.17 | 90 | 91 | 91 |
| linuxrwlocks(8) | 6984234 | 311.93 | 1176.13 | 1119.23 | 90 | 91 | 91 |
| filesystem(22) | 512 | 0.72 | 0.62 | 0.34 | 86 | 84 | 84 |
| filesystem(24) | 2048 | 2.84 | 2.97 | 1.32 | 86 | 187 | 84 |
| filesystem(26) | 8192 | 11.88 | 15.71 | 5.66 | 85 | 622 | 84 |
| indexer(15) | 4096 | 11.07 | 8.58 | 5.65 | 89 | 116 | 90 |
| indexer(16) | 32768 | 90.14 | 80.37 | 46.46 | 89 | 464 | 90 |
| indexer(17) | 262144 | 736.78 | 827.02 | 399.87 | 89 | 3030 | 90 |
| lastzero(10) | 3328 | 0.07 | 0.34 | 0.27 | 85 | 84 | 84 |
| lastzero(15) | 147456 | 3.19 | 24.46 | 15.09 | 85 | 276 | 84 |
| lastzero(20) | 6029312 | 152.13 | 1828.92 | 786.19 | 85 | 8883 | 84 |
| exp-mem3(7) | 10080 | 0.22 | 0.67 | 0.54 | 86 | 104 | 85 |
| exp-mem3(8) | 80640 | 1.96 | 6.15 | 4.61 | 86 | 506 | 85 |
| exp-mem3(9) | 725760 | 19.11 | 73.68 | 44.83 | 86 | 4489 | 85 |
| dispatcher(4) | 6854 | 1.15 | 1.75 | 1.47 | 90 | 90 | 90 |
| dispatcher(5) | 151032 | 34.66 | 55.07 | 42.76 | 89 | 407 | 90 |
| dispatcher(6) | 4057388 | 1245.13 | 2333.51 | 1424.57 | 89 | 9097 | 90 |
| poke(10) | 135944 | 88.54 | 96.30 | 63.45 | 90 | 791 | 90 |
| poke(15) | 728559 | 874.76 | 891.26 | 479.03 | 89 | 5527 | 90 |
| poke(20) | 2366924 | 4502.45 | 4356.59 | 2008.92 | 90 | 22383 | 90 |
| length-param(2,1024) | 4 | 0.14 | 0.05 | 0.06 | 85 | 84 | 84 |
| length-param(2,8196) | 4 | 7.95 | 0.16 | 0.14 | 95 | 101 | 89 |
| length-param(2,65536) | 4 | 1413.00 | 1.13 | 0.90 | 389 | 441 | 343 |

The first group consists of three programs (circular-buffer from SCTBench [47], fib-bench from SV-Comp [45], and the linuxrwlocks from SATCheck [15]). Here, all algorithms consume memory that stays constant as the size of the program and the number of executions explored increase. We can therefore compare the raw performance of the implementation of these three DPOR algorithms. POP's implementation is fastest on circular-buffer, while TruSt's is fastest on the two other programs. However, notice that all three implementations scale similarly.

The second group consists of the two benchmarks (filesystem and indexer) from the "classic" DPOR paper of Flanagan and Godefroid [17]. Here, Optimal DPOR shows an increase in memory consumption (measured in MB), while the other two algorithms use constant memory. POP is fastest here by approximately 2×.

The third group, consisting of lastzero [1] and exp-mem3,[4] two synthetic benchmarks also used in the TruSt paper [29, Table 1], shows a similar picture in terms of memory consumption: Optimal DPOR's increases more noticeably here, while the two other algorithms use memory that stays constant. Time-wise, TruSt is 2–5× faster than POP, which in turn is 2× faster than Optimal.

The fourth group, consisting of two concurrent data structure programs (dispatcher and poke) from the Quasi-Optimal POR paper [40], shows Optimal's memory explosion more profoundly, and provides further evidence of the good memory performance of the TruSt and POP algorithms. Time-wise, there is no clear winner here, with TruSt's implementation being a bit faster on dispatcher, and with POP's being faster and scaling slightly better than TruSt's on poke.

Finally, let us examine the algorithms' performance on length-param($T$,$N$), a synthetic but simple program in which a number of threads (just two here) issue $N$ stores and loads to thread-specific global variables, followed by a store and a load to a variable shared between threads. The total number of executions is just four here, but the executions grow in length. One can clearly see the superior time performance of sequence-based DPOR algorithms, such as Optimal and POP, compared to TruSt's graph-based algorithm that needs to perform consistency checks for the executions it constructs. As can be seen, these checks can become quite expensive (esp. if their implementation has sub-optimal complexity, as it is probably the case here). In contrast, sequence-based DPOR algorithms naturally generate consistent executions (for memory models such as SC). We can also notice that POP performs slightly better than Optimal in terms of memory.

As further evidence for POP's implementation being quite robust, we mention in passing that we also conducted an experiment in which we used POP to check, under the SC memory model, the code of Linux kernel's Hierarchical Read-Copy Update (Tree RCU) implementation using the code repository of a conference paper by Kokologiannakis and Sagonas [32]. Since GenMC cannot handle RCU's code, we do not include these results in the table.

Wrapping up our evaluation, we can make the following two general claims:

1. Both POP and TruSt live up to their promise about performing SMC exploration which is optimal (w.r.t. the Mazurkiewicz equivalence) but also with polynomial (in fact, in practice, constant) space consumption.
2. The implementation of the POP algorithm consistently outperforms that of Optimal DPOR in Nidhugg. This is mostly due to increased simplicity.

## 7  Related Work

Since its introduction in the tools Verisoft [19,20] and CHESS [39], stateless model checking has been an important technique for analyzing correctness of concurrent

---

[4] exp-mem3 is slight variant of the exp-mem program used in the TruSt paper. It uses atomic stores and loads instead of fetch-and-adds (FAAs), because the current implementation of Optimal DPOR (and POP) in Nidhugg employs an optimization which treats independent FAAs as non-conflicting [25] and explores only one trace on the exp-mem program independently of the benchmark's parameter.

programs. Dynamic partial order reduction [17, 44] has enabled a significantly increased efficiency for covering all interleavings, which has been adapted to many different settings and computational models, including actor programs [46], abstract computational models [27], event driven programs [4, 24, 35], and MPI programs [42]. DPOR has been adapted for weak memory models including TSO [2, 15, 48], Release-Acquire [7], POWER [6], and C11 [28], and also been applied to real life programs [32]. DPOR has been extended with features for efficiently handling spinloops and blocking constructs [25, 31],

An important advancement has been the introduction of *optimal* DPOR algorithms, which guarantee to explore *exactly* one execution from each equivalence class [1], and therefore achieve exponential-time reduction over non-optimal algorithms. This saving came at the cost of worst-case exponential (in the size of the program) memory consumption [3]. The strive for covering the space of all interleavings with fewer representative executions inspired DPOR algorithms for even weaker equivalences than Mazurkiewicz trace equivalence, such as equivalence based on observers [10], reads-from equivalence [5, 12, 13], conditional independence [9], context-sensitive independence and observers [8], or on the maximal causal model [23]. These approaches explore fewer traces than approaches based on Mazurkiewicz trace equivalence at the cost of potentially expensive (often NP-hard) consistency checks. Another line of work uses unfoldings [37] to further reduce the number of interleavings that must be considered [26, 40, 43]; these techniques incur significantly larger cost per test execution than the previously mentioned ones.

DPOR has also been adapted for weak memory models using an approach in which executions are represented as graphs, where nodes represent read and write operations, and edges represent reads-from and coherence relations; this allows the algorithm to be parametric on a specific memory model, at the cost of calling a memory-model oracle [28, 30, 33]. For this graph-based setting, an optimal DPOR algorithm with worst-case polynomial space consumption, called TruSt, was recently presented [29]. POP is also optimal with worst-case polynomial space consumption. Since it is designed for a sequence-based representation of executions, POP must be designed differently. In analogy with the parsimonious race reversal technique, TruSt has a technique for reversing each race only once, which is based on a *maximal extension* criterion. POP adapts TruSt's strategy of eager race reversal to avoid potentially space-consuming accumulation of schedules. Finally, since TruSt operates in a graph-based setting, it reverses write-read races by changing the source of a read-from relation in the graph, instead of constructing a new schedule. Therefore redundant exploration of read-schedules is prevented by careful book-keeping instead of using sleep sets, which POP represents in a compact parsimonious way. The experimental results show that TruSt and POP have comparable performance for small and modest-size programs, but that POP is superior for programs with long executions, since the graph-based approach has difficulties to scale for long executions.

An alternative to DPOR for limiting the number of explored executions is to cover only a subset of all executions. Various heuristics for choosing this subset have been developed, including delay bounding [16], preemption bounding [38], and probabilistic strategies [11]. Such techniques can be effective in finding bugs in concurrent programs, but not prove their absence.

## 8  Conclusion

In this paper, we have presented POP, a new optimal DPOR algorithm for analyzing multi-threaded programs under SC. POP combines several novel algorithmic techniques, which allow efficiency improvements over previous such DPOR algorithms, both in time and space. In particular, its space consumption is polynomial in the size of the analyzed program. Our experiments on a wide variety of benchmarks show that POP always outperforms Optimal DPOR, the state-of-the-art sequence-based optimal DPOR algorithm, and offers performance comparable with TruSt, the state-of-the-art graph-based DPOR algorithm. Moreover, by being sequence-based, its implementation scales much better than TruSt's on programs with long executions.

As future work, it would be interesting to investigate the effect of applying POP's novel algorithmic techniques on DPOR algorithms tailored for different computational models, and for analyzing programs under weak concurrency memory models such as TSO and PSO.

**Acknowledgements** This research was partially funded by research grants from the Swedish Research Council (Vetenskapsrådet) and from the Swedish Foundation for Strategic Research through project aSSIsT. We thank these funding agencies and the anonymous CAV 2024 reviewers for their comments.

## References

1. Abdulla, P., Aronis, S., Jonsson, B., Sagonas, K.: Optimal dynamic partial order reduction. In: Symposium on Principles of Programming Languages. pp. 373–384. POPL 2014, ACM, New York, NY, USA (2014). https://doi.org/10.1145/2535838.2535845, http://doi.acm.org/10.1145/2535838.2535845
2. Abdulla, P.A., Aronis, S., Atig, M.F., Jonsson, B., Leonardsson, C., Sagonas, K.: Stateless model checking for TSO and PSO. In: Tools and Algorithms for the Construction and Analysis of Systems. LNCS, vol. 9035, pp. 353–367. Springer, Berlin, Heidelberg (2015). https://doi.org/10.1007/978-3-662-46681-0_28, http://dx.doi.org/10.1007/978-3-662-46681-0_28
3. Abdulla, P.A., Aronis, S., Jonsson, B., Sagonas, K.: Source sets: A foundation for optimal dynamic partial order reduction. Journal of the ACM **64**(4), 25:1–25:49 (Sep 2017). https://doi.org/10.1145/3073408, http://doi.acm.org/10.1145/3073408
4. Abdulla, P.A., Atig, M.F., Bønneland, F.M., Das, S., Jonsson, B., Lång, M., Sagonas, K.: Tailoring stateless model checking for event-driven multi-threaded programs. In: André, É., Sun, J. (eds.) Automated Technology for Verification and Analysis - 21st International Symposium, ATVA 2023, Proceedings, Part II. LNCS, vol. 14216, pp. 176–198. Springer (Oct 2023). https://doi.org/10.1007/978-3-031-45332-8_9, https://doi.org/10.1007/978-3-031-45332-8_9

5. Abdulla, P.A., Atig, M.F., Jonsson, B., Lång, M., Ngo, T.P., Sagonas, K.: Optimal stateless model checking for reads-from equivalence under sequential consistency. Proc. ACM Program. Lang. **3**(OOPSLA), 150:1–150:29 (Oct 2019). https://doi.org/10.1145/3360576, https://doi.org/10.1145/3360576
6. Abdulla, P.A., Atig, M.F., Jonsson, B., Leonardsson, C.: Stateless model checking for POWER. In: Computer Aided Verification. LNCS, vol. 9780, pp. 134–156. Springer International Publishing, Cham (2016). https://doi.org/10.1007/978-3-319-41540-6_8, http://dx.doi.org/10.1007/978-3-319-41540-6_8
7. Abdulla, P.A., Atig, M.F., Jonsson, B., Ngo, T.P.: Optimal stateless model checking under the release-acquire semantics. Proc. ACM on Program. Lang. **2**(OOPSLA), 135:1–135:29 (2018). https://doi.org/10.1145/3276505, http://doi.acm.org/10.1145/3276505
8. Albert, E., de la Banda, M.G., Gómez-Zamalloa, M., Isabel, M., Stuckey, P.J.: Optimal dynamic partial order reduction with context-sensitive independence and observers. J. Syst. Softw. **202**, 111730 (2023). https://doi.org/10.1016/J.JSS.2023.111730, https://doi.org/10.1016/j.jss.2023.111730
9. Albert, E., Gómez-Zamalloa, M., Isabel, M., Rubio, A.: Constrained dynamic partial order reduction. In: Computer Aided Verification. LNCS, vol. 10982, pp. 392–410. Springer, Cham (Jul 2018). https://doi.org/10.1007/978-3-319-96142-2_24, https://doi.org/10.1007/978-3-319-96142-2_24
10. Aronis, S., Jonsson, B., Lång, M., Sagonas, K.: Optimal dynamic partial order reduction with observers. In: Tools and Algorithms for the Construction and Analysis of Systems - 24th International Conference. LNCS, vol. 10806, pp. 229–248. Springer, Cham (Apr 2018). https://doi.org/10.1007/978-3-319-89963-3_14, https://doi.org/10.1007/978-3-319-89963-3_14
11. Burckhardt, S., Kothari, P., Musuvathi, M., Nagarakatte, S.: A randomized scheduler with probabilistic guarantees of finding bugs. In: Proceedings of the Fifteenth Edition of ASPLOS on Architectural Support for Programming Languages and Operating Systems. pp. 167–178. ASPLOS XV, ACM, New York, NY, USA (2010). https://doi.org/10.1145/1736020.1736040, http://doi.acm.org/10.1145/1736020.1736040
12. Chalupa, M., Chatterjee, K., Pavlogiannis, A., Sinha, N., Vaidya, K.: Data-centric dynamic partial order reduction. Proc. ACM on Program. Lang. **2**(POPL), 31:1–31:30 (Jan 2018). https://doi.org/10.1145/3158119, http://doi.acm.org/10.1145/3158119
13. Chatterjee, K., Pavlogiannis, A., Toman, V.: Value-centric dynamic partial order reduction. Proc. ACM Program. Lang. **3**(OOPSLA), 124:1–124:29 (Oct 2019). https://doi.org/10.1145/3360550, https://doi.org/10.1145/3360550
14. Christakis, M., Gotovos, A., Sagonas, K.: Systematic testing for detecting concurrency errors in Erlang programs. In: Sixth IEEE International Conference on Software Testing, Verification and Validation. pp. 154–163. ICST 2013, IEEE, Los Alamitos, CA, USA (Mar 2013). https://doi.org/10.1109/ICST.2013.50, https://doi.org/10.1109/ICST.2013.50
15. Demsky, B., Lam, P.: SATCheck: SAT-directed stateless model checking for SC and TSO. In: Proceedings of the 2015 ACM SIGPLAN International Conference on Object-Oriented Programming, Systems, Languages, and Applications. pp. 20–36. OOPSLA 2015, ACM, New York, NY, USA (2015). https://doi.org/10.1145/2814270.2814297, http://doi.acm.org/10.1145/2814270.2814297
16. Emmi, M., Qadeer, S., Rakamaric, Z.: Delay-bounded scheduling. In: Proceedings of the 38th ACM SIGPLAN-SIGACT Symposium on Principles of Programming Languages. pp. 411–422. POPL 2011, ACM (Jan 2011). https://doi.org/10.1145/1926385.1926432, https://doi.org/10.1145/1926385.1926432

17. Flanagan, C., Godefroid, P.: Dynamic partial-order reduction for model checking software. In: Principles of Programming Languages, (POPL). pp. 110–121. ACM, New York, NY, USA (Jan 2005). https://doi.org/10.1145/1040305.1040315, http://doi.acm.org/10.1145/1040305.1040315
18. Godefroid, P.: Partial-Order Methods for the Verification of Concurrent Systems: An Approach to the State-Explosion Problem. Ph.D. thesis, University of Liège (1996). https://doi.org/10.1007/3-540-60761-7, http://www.springer.com/gp/book/9783540607618, also, volume 1032 of LNCS, Springer.
19. Godefroid, P.: Model checking for programming languages using VeriSoft. In: Principles of Programming Languages, (POPL). pp. 174–186. ACM Press, New York, NY, USA (Jan 1997). https://doi.org/10.1145/263699.263717, http://doi.acm.org/10.1145/263699.263717
20. Godefroid, P.: Software model checking: The VeriSoft approach. Formal Methods in System Design **26**(2), 77–101 (Mar 2005). https://doi.org/10.1007/s10703-005-1489-x, http://dx.doi.org/10.1007/s10703-005-1489-x
21. Godefroid, P., Hanmer, R.S., Jagadeesan, L.: Model checking without a model: An analysis of the heart-beat monitor of a telephone switch using VeriSoft. In: Proceedings of the ACM SIGSOFT International Symposium on Software Testing and Analysis. pp. 124–133. ISSTA, ACM, New York, NY, USA (Mar 1998). https://doi.org/10.1145/271771.271800, https://doi.org/10.1145/271771.271800
22. Godefroid, P., Holzmann, G.J., Pirottin, D.: State-space caching revisited. Formal Methods in System Design **7**(3), 227–241 (1995). https://doi.org/10.1007/BF01384077, http://dx.doi.org/10.1007/BF01384077
23. Huang, J.: Stateless model checking concurrent programs with maximal causality reduction. In: Proceedings of the 36th ACM SIGPLAN Conference on Programming Language Design and Implementation. pp. 165–174. PLDI 2015, ACM, New York, NY, USA (2015). https://doi.org/10.1145/2737924.2737975, http://doi.acm.org/10.1145/2737924.2737975
24. Jensen, C.S., Møller, A., Raychev, V., Dimitrov, D., Vechev, M.T.: Stateless model checking of event-driven applications. In: Proceedings of the 2015 ACM SIGPLAN International Conference on Object-Oriented Programming, Systems, Languages, and Applications. pp. 57–73. OOPSLA 2015, ACM, New York, NY, USA (2015). https://doi.org/10.1145/2814270.2814282, https://doi.org/10.1145/2814270.2814282
25. Jonsson, B., Lång, M., Sagonas, K.: Awaiting for Godot: Stateless model checking that avoids executions where nothing happens. In: Griggio, A., Rungta, N. (eds.) 22nd Formal Methods in Computer-Aided Design. pp. 284–293. FMCAD 2022, IEEE (Oct 2022). https://doi.org/10.34727/2022/ISBN.978-3-85448-053-2_35, https://doi.org/10.34727/2022/isbn.978-3-85448-053-2_35
26. Kähkönen, K., Saarikivi, O., Heljanko, K.: Using unfoldings in automated testing of multithreaded programs. In: IEEE/ACM International Conference on Automated Software Engineering. pp. 150–159. ASE'12, ACM, New York, NY, USA (2012). https://doi.org/10.1145/2351676.2351698, http://dl.acm.org/citation.cfm?id=2351676
27. Kastenberg, H., Rensink, A.: Dynamic partial order reduction using probe sets. In: van Breugel, F., Chechnik, M. (eds.) Concurrency Theory. LNCS, vol. 5201, pp. 233–247. Springer (2008). https://doi.org/10.1007/978-3-540-85361-9_21
28. Kokologiannakis, M., Lahav, O., Sagonas, K., Vafeiadis, V.: Effective stateless model checking for C/C++ concurrency. Proc. ACM on Program. Lang. **2**(POPL), 17:1–17:32 (Jan 2018). https://doi.org/10.1145/3158105, https://doi.org/10.1145/3158105

29. Kokologiannakis, M., Marmanis, I., Gladstein, V., Vafeiadis, V.: Truly stateless, optimal dynamic partial order reduction. Proc. ACM Program. Lang. **6**(POPL), 1–28 (2022). https://doi.org/10.1145/3498711, https://doi.org/10.1145/3498711
30. Kokologiannakis, M., Raad, A., Vafeiadis, V.: Model checking for weakly consistent libraries. In: Proceedings of the 40th ACM SIGPLAN Conference on Programming Language Design and Implementation. pp. 96–110. PLDI 2019, ACM, New York, NY, USA (Jun 2019). https://doi.org/10.1145/3314221.3314609, https://doi.org/10.1145/3314221.3314609
31. Kokologiannakis, M., Ren, X., Vafeiadis, V.: Dynamic partial order reductions for spinloops. In: Formal Methods in Computer Aided Design. pp. 163–172. FMCAD 2021, IEEE (Oct 2021). https://doi.org/10.34727/2021/isbn.978-3-85448-046-4_25, https://doi.org/10.34727/2021/isbn.978-3-85448-046-4_25
32. Kokologiannakis, M., Sagonas, K.: Stateless model checking of the Linux kernel's read–copy update (RCU). Software Tools for Technology Transfer **21**(3), 287–306 (Jun 2019). https://doi.org/10.1007/s10009-019-00514-6, https://doi.org/10.1007/s10009-019-00514-6
33. Kokologiannakis, M., Vafeiadis, V.: HMC: model checking for hardware memory models. In: Larus, J.R., Ceze, L., Strauss, K. (eds.) Architectural Support for Programming Languages and Operating Systems. pp. 1157–1171. ASPLOS '20, ACM (Mar 2020). https://doi.org/10.1145/3373376.3378480, https://doi.org/10.1145/3373376.3378480
34. Kokologiannakis, M., Vafeiadis, V.: GenMC: A model checker for weak memory models. In: Computer Aided Verification - 33rd International Conference, CAV 2021, Proceedings, Part I. LNCS, vol. 12759, pp. 427–440. Springer (Jul 2021). https://doi.org/10.1007/978-3-030-81685-8_20, https://doi.org/10.1007/978-3-030-81685-8_20
35. Maiya, P., Gupta, R., Kanade, A., Majumdar, R.: Partial order reduction for event-driven multi-threaded programs. In: Chechik, M., Raskin, J. (eds.) Tools and Algorithms for the Construction and Analysis of Systems (TACAS 2016). LNCS, vol. 9636, pp. 680–697. Springer, Berlin, Heidelberg (Apr 2016). https://doi.org/10.1007/978-3-662-49674-9_44, https://doi.org/10.1007/978-3-662-49674-9_44
36. Mazurkiewicz, A.: Trace theory. In: Brauer, W., Reisig, W., Rozenberg, G. (eds.) Petri Nets: Applications and Relationships to Other Models of Concurrency. LNCS, vol. 255, pp. 279–324. Springer, Berlin Heidelberg (1987). https://doi.org/10.1007/3-540-17906-2_30, http://dx.doi.org/10.1007/3-540-17906-2_30
37. McMillan, K.L.: A technique of a state space search based on unfolding. Formal Methods in System Design **6**(1), 45–65 (1995). https://doi.org/10.1007/BF01384314, https://doi.org/10.1007/BF01384314
38. Musuvathi, M., Qadeer, S.: Iterative context bounding for systematic testing of multithreaded programs. In: Proceedings of the ACM SIGPLAN 2007 Conference on Programming Language Design and Implementation. pp. 446–455. PLDI 2007, ACM (Jun 2007). https://doi.org/10.1145/1250734.1250785, https://doi.org/10.1145/1250734.1250785
39. Musuvathi, M., Qadeer, S., Ball, T., Basler, G., Nainar, P.A., Neamtiu, I.: Finding and reproducing heisenbugs in concurrent programs. In: Proceedings of the 8th USENIX Symposium on Operating Systems Design and Implementation. pp. 267–280. OSDI '08, USENIX Association, Berkeley, CA, USA (Dec 2008), http://dl.acm.org/citation.cfm?id=1855741.1855760
40. Nguyen, H.T.T., Rodríguez, C., Sousa, M., Coti, C., Petrucci, L.: Quasi-optimal partial order reduction. In: Chockler, H., Weissenbacher, G. (eds.) Computer Aided Verification. LNCS, vol. 10982, pp. 354–371. Springer, Cham (2018). https://doi.org/10.1007/978-3-319-96142-2_22

41. Norris, B., Demsky, B.: A practical approach for model checking C/C++11 code. ACM Trans. Program. Lang. Syst. **38**(3), 10:1–10:51 (May 2016). https://doi.org/10.1145/2806886, http://doi.acm.org/10.1145/2806886
42. Palmer, R., Gopalakrishnan, G., Kirby, R.M.: Semantics driven dynamic partial-order reduction of MPI-based parallel programs. In: Ur, S., Farchi, E. (eds.) Proceedings of the 5th Workshop on Parallel and Distributed Systems: Testing, Analysis, and Debugging. pp. 43–53. PADTAD 2007, ACM (Jul 2007). https://doi.org/10.1145/1273647.1273657, https://doi.org/10.1145/1273647.1273657
43. Rodríguez, C., Sousa, M., Sharma, S., Kroening, D.: Unfolding-based partial order reduction. In: 26th International Conference on Concurrency Theory (CONCUR 2015). LIPIcs, vol. 42, pp. 456–469. Schloss Dagstuhl–Leibniz-Zentrum fuer Informatik (Aug 2015). https://doi.org/10.4230/LIPIcs.CONCUR.2015.456, http://dx.doi.org/10.4230/LIPIcs.CONCUR.2015.456
44. Sen, K., Agha, G.: A race-detection and flipping algorithm for automated testing of multi-threaded programs. In: Hardware and Software, Verification and Testing. LNCS, vol. 4383, pp. 166–182. Springer, Berlin Heidelberg (2007). https://doi.org/10.1007/978-3-540-70889-6_13, https://doi.org/10.1007/978-3-540-70889-6_13
45. SV-COMP: Competition on Software Verification. https://sv-comp.sosy-lab.org/2019 (2019), [Online; accessed 2019-03-24]
46. Tasharofi, S., Karmani, R.K., Lauterburg, S., Legay, A., Marinov, D., Agha, G.: TransDPOR: A novel dynamic partial-order reduction technique for testing actor programs. In: Giese, H., Rosu, G. (eds.) Formal Techniques for Distributed Systems. LNCS, vol. 7273, pp. 219–234. Springer, Berlin Heidelberg (2012). https://doi.org/10.1007/978-3-642-30793-5_14
47. Thomson, P., Donaldson, A.F., Betts, A.: Concurrency testing using controlled schedulers: An empirical study. ACM Trans. Parallel Comput. **2**(4), 23:1–23:37 (2016). https://doi.org/10.1145/2858651, http://doi.acm.org/10.1145/2858651
48. Zhang, N., Kusano, M., Wang, C.: Dynamic partial order reduction for relaxed memory models. In: Programming Language Design and Implementation (PLDI). pp. 250–259. ACM, New York, NY, USA (Jun 2015). https://doi.org/10.1145/2737924.2737956, http://doi.acm.org/10.1145/2737924.2737956

## A  Proofs of Correctness

In this appendix, we provide the proofs for Theorem 1 and 2.

**Theorem 1.** *For a terminating program $P$, the POP algorithm has the properties that (i) for each maximal execution $E$ of $P$, it explores some execution $E'$ with $E' \simeq E$, (ii) it never explores two different but equivalent maximal executions, and (iii) it is never blocked (at line 13) unless the execution currently exploring is a maximal one.*

We will establish the three properties of Theorem 1 separately. Property (i) follows from Theorem 3 and Lemma 2, property (ii) is established in Theorem 4, and property (iii) is established in Theorem 5. Let us first introduce some concepts. For a sequence $w$, we say that a read event $e$ accessing variable $x$ occurring in $w$ is a *first read on $x$ in $w$* if there is no other event $e'$ occurring in $w$ accessing $x$ such that $e' \xrightarrow{\text{hb}} e$. For an execution $E \cdot w$, we say that the call $\texttt{UpdSeq}(w, \mathit{SSChar}[E])$ of Algorithm 1 is *blocking* if it returns *block*, and *not blocking* otherwise. The set of schedules characterized by the sleep set expression $\psi$ is denoted by $[\![\psi]\!]$. Similarly, $[\![SSChar[E]]\!]$ denotes the union of the sets $[\![\psi]\!]$, where $\psi$ is in $SSChar[E]$. Next we state and prove Lemma 2, which shows the correctness of sleep set manipulation function $\texttt{UpdSeq}(w, SSChar)$.

**Lemma 2.** *For a call* $\texttt{Explore}(E)$ *of Algorithm 1 and an execution sequence* $E \cdot w$, *the call* $\texttt{UpdSeq}(w, SSChar[E])$ *is blocking iff there is a read schedule* $\sigma$ *with* $\sigma \sqsubseteq w$ *such that* $\sigma \in [\![SSChar[E]]\!]$.

*Proof.* It is sufficient to prove that for a sleep set expression $\psi$ in $SSChar[E]$, $\sigma$ is in $[\![\psi]\!]$ iff $\texttt{UpdSeq}(w, SSChar[E])$ is blocking. There are two cases:

1. $\psi$ is of the form $P_m \triangleright x^R$, which matches the argument of the $\texttt{UpdSE}$ function at line 12. Now, we prove the forward direction assuming $\sigma$ is in $[\![\psi]\!]$. As no write event accesses $x$ in $\sigma$, updated $\psi$ is not filtered out in line 11. Hence, for each event $e$ from the beginning to the end of $\sigma$, the algorithm calls $\texttt{UpdSE}(e, \psi^\sigma)$ at line 12, which immediately calls $\texttt{UpdP}$ until blocking, where $\psi^\sigma$ is $\psi$ initially and updates at each call. There are two cases corresponding to the cases (A) and (B) of Proposition 1.
   Case (A) (contained schedule): Because of condition (i) of case (A), the condition at line 35 is always satisfied, and line 40 is executed except for the last call because only the head of $\sigma$ is a read event on $x$. For the last call of $\texttt{UpdP}$ with the head of $\sigma$, the conditions at line 37 are satisfied because the head of $\sigma$ is a read event accessing $x$, and condition (ii) of case (A) is true. Also, the condition at line 38 is true because no other read event on $x$ is in $\sigma$, and the function $\texttt{UpdP}$ returns *block*.
   Case (B) (conflicting schedule): Because of condition (i) of case (B), in one or more calls, including the last one of $\texttt{UpdP}$, condition at line 34 is satisfied, and $e$ is added to some $C_l$. The condition at line 35 is satisfied in other calls. The algorithm executes line 40 in every call except the last one because the head of the schedule is the only read event accessing $x$. In the last call of $\texttt{UpdP}$ with the head of $\sigma$, the condition at line 37 is satisfied because the head of $\sigma$ is a read event accessing $x$, and condition (ii) of case (B) is true. Also, the condition at line 38 is satisfied because no other read event on $x$ is in $\sigma$, and the function $\texttt{UpdP}$ returns *block*.
   Hence, the forward direction is established. As each proof step is reversible, the other direction also holds.
2. otherwise, $\psi$ is of the form $P[\psi']^D \triangleright x^R$, representing inherited schedule, matches the argument of the $\texttt{UpdSE}$ function at line 14. Here, case (C) of Proposition 1 applies since $\psi$ characterizes inherited schedules. Case (C) implies that there is a read-schedule $\sigma_i$ such that $\sigma_i \sqsubseteq \sigma$ and $\sigma_i' \propto \sigma_i$, where $\sigma_i'$ is characterized by $\psi'$. There are two cases.
   Case 1) $hd(\sigma_i).var \neq x$. Condition (i) implies that there are three types of events in $(\sigma \setminus hd(\sigma))$. The first type matches the prefix $P$. The calls of $\texttt{UpdSE}$ with these events are not blocking as they call $\texttt{UpdP}$ at line 28, and the condition at line 35 is satisfied, thereafter executing line 40. The second type matches $\sigma_i'$. The calls of $\texttt{UpdSE}$ with these events will call

UpdSE recursively with $\psi'$ at line 21. For all the events of this kind except $hd(\sigma'_i)$, the updated expression will be returned at line 27. Assuming $\sigma'_i$ is in $[\![\psi']\!]$, the updated expression returned by recursive UpdSE call with $hd(\sigma_i)$ will have $hd(\sigma_i)$ in the set $D$. The last and remaining type of event is $hd(\sigma)$. Calling UpdSE with $hd(\sigma)$ will satisfy the condition at line 17, and the algorithm returns *block* at line 19 as no other read event on $x$ occurs in $\sigma$.

- Case 2) $hd(\sigma_i).var = x$. The proof in this case is almost the same as the previous case. The only difference is $hd(\sigma)$ also falls in the second type, meaning the call of $\texttt{UpdSE}(hd(\sigma), \psi')$ at line 21 will return *block*. Because of condition (ii) and (iii) of this case in Proposition 1, the condition at line 23 is true, meaning the function call returns *block* at line 25. Condition (ii) of this case tells that no other read event on $x$ except the $hd(\sigma)$ occurs in $\sigma$.

We established the forward direction. Now we can establish the other direction using the following steps. We assume that there is a sequence of calls to the UpdSE function at line 14 with the events occurring in $\sigma$ from beginning to the end where the last call returns *block*. The last call with the event $hd(\sigma)$ returns *block* at either line 19 or line 25. When the algorithm returns block at line 19 then case (C) (1) of Proposition 1 is satisfied. When the algorithm returns block at line 25 then case (C) (2) of Proposition 1 is satisfied.

□

**Theorem 3 (Correctness of POP).** *Whenever a call to* Explore$(E)$ *returns during Algorithm 1, then for each maximal executions $E \cdot w$ such that* UpdSeq$(w, \textit{SSChar}[E])$ *is not blocking, the call* Explore$(E)$ *has explored an execution in* $[E \cdot w]_{\simeq}$.

Since the initial call to Explore is Explore$(\langle\rangle)$ with $\textit{SSChar}[\langle\rangle]$ empty, Theorem 3 implies that for all maximal executions $E$ the algorithm explores some execution in $[E]_{\simeq}$, thereby establishing property (i) of Theorem 1.

The proof of Theorem 3 in its turn relies on the following key lemma

**Lemma 3.** *Whenever a call to* Explore$(E)$ *returns during Algorithm 1, then for all schedules $w \cdot e$ such that* UpdSeq$((w \cdot e), \textit{SSChar}[E])$ *is not blocking, the call* Explore$(E)$ *will explore* Explore$(E \cdot w')$ *for some sequence $w'$ containing the event $e$ such that (i) $w \cdot e \simeq e \downarrow^{w'}$ and (ii) $e$ is fresh in $w'$ after $E$.*

*Proof.* We prove this by induction on the length of longest maximal execution after $E$.

*Base Case:* When $E$ is a maximal execution the lemma trivially holds as there is no $w \cdot e$ to explore after $E$.

*Inductive Step:* Let $e'$ be the first event such that a call of form Explore$(E \cdot e')$ is performed (at line 15). By the test at line 13, UpdSeq$(\langle e'\rangle, \textit{SSChar}[E]) \neq block$. If $e' = e$ (implying $w = \langle\rangle$), then the lemma is established directly. Otherwise, we consider two cases:

**Case 1:** $e' \sim w \cdot e$. Let $u \cdot e$ be $(w \cdot e) \backslash e'$. In this case, $E \cdot e' \cdot u \cdot e$ is an execution sequence. For each event $e_w$ other than $e'$ occurring in $w$, it is true that $e_w \xrightarrow{\text{hb}}_{w \cdot e} e$ but not $e_w \xrightarrow{\text{hb}}_{w \cdot e} e'$, which implies that $e_w \xrightarrow{\text{hb}}_{u \cdot e} e'$, i.e., $u \cdot e$ is a schedule. We will apply the I.H. using $E \cdot e'$ in place of $E$ and $u \cdot e$ in place of $w \cdot e$ after showing $\texttt{UpdSeq}((u \cdot e), \textit{SSChar}[E \cdot e'])$ is not blocking. The premise of this lemma says implies that the call $\texttt{UpdSeq}((e' \cdot u \cdot e), \textit{SSChar}[E])$ is not blocking. The function call will satisfy exactly one of the three cases of the $\texttt{UpdSeq}$ function definition. The first case in line 8 will not be satisfied because $e' \cdot u \cdot e$ is not an empty sequence. The second case in line 9 will not be satisfied because the test at line 13 of Algorithm 1 succeeds. The third case in line 11 satisfies implying the call $\texttt{UpdSeq}((u \cdot e), \textit{SSChar}[E \cdot e'])$ is not blocking. Hence, the I.H. implies that the call $\texttt{Explore}(E \cdot e')$ will explore $\texttt{Explore}(E \cdot e' \cdot w'')$ for some sequence $w''$ such that (i) $(u \cdot e) \simeq e \downarrow^{w''}$, and (ii) $e$ is fresh in $w''$ after $E \cdot e'$. We then derive properties (i) and (ii) in the lemma as follows. Property (i) is straightforward. Property (ii) follows by the fact that $e'$ is not in a schedule since it is the first event explored after $E$.

**Case 2:** $e' \not\sim w \cdot e$. Considering $e'$ accesses $x$, there another conflicting event accessing $x$ must occur in $w \cdot e$. If the first access to $x$ in $w \cdot e$ is a write, then let $e_u$ be this write event; otherwise, let the event $e_u$ range over the set of first reads on $x$ in $w \cdot e$. Since $e_u \downarrow^{w \cdot e}$ is a schedule by construction, it does not contain $e'$. Let $u \cdot e_u$ be $e_u \downarrow^{w \cdot e}$. It follows that $E \cdot e' \cdot u \cdot e_u$ is an execution sequence for each such $e_u$. Using $E \cdot e'$ in place of $E$ and $u \cdot e_u$ in place of $w \cdot e$, we can apply the I.H. after establishing $\texttt{UpdSeq}((u \cdot e_u), \textit{SSChar}[E \cdot e'])$ is not blocking. Because $u \cdot e_u \sqsubseteq w \cdot e$, it is true that $\texttt{UpdSeq}((u \cdot e_u), \textit{SSChar}[E])$ is not blocking. The test at line 13 stating $\texttt{UpdSeq}(e', \textit{SSChar}[E])$, succeeds returning $\textit{SSChar}[E \cdot e']$ at line 14. The last two statements imply that $\texttt{UpdSeq}((u \cdot e_u), \textit{SSChar}[E \cdot e'])$ is blocking only when there is a read event $e'_u$ occurring in $u \cdot e_u$, such that $e' \xrightarrow{\text{hb}} e'_u$ (line 17), which will never occur because $e'$ will be a write event removing the sleep set expression (line 11). Hence we establish $\texttt{UpdSeq}((u \cdot e_u), \textit{SSChar}[E \cdot e'])$ is not blocking.

We can thus apply the I.H., which implies that the call $\texttt{Explore}(E \cdot e')$ will explore $\texttt{Explore}(E \cdot e' \cdot w')$ for some sequence $w'$ such that (i) $(u \cdot e_u) \simeq e_u \downarrow^{w'}$, and (ii) $e_u$ is fresh in $w'$ after $E \cdot e'$. Thus the race $e' \lesssim_{E \cdot e' \cdot w'} e_u$ is parsimonious, and the algorithm creates a schedule $\sigma_u = e_u \downarrow^{w'}$ for each $e_u$. Consider the smallest schedule $\sigma$ among $\sigma_u$s in the $\propto$ order. It gives rise to a call $\texttt{Explore}(E \cdot \sigma)$ at line line 12. If $w \cdot e = \sigma$, the proof is done. Otherwise, let $w_\sigma \cdot e$ be $(w \cdot e) \backslash \sigma$. Then $E \cdot \sigma \cdot w_\sigma \cdot e$ is an execution sequence. We need to show that $w_\sigma \cdot e$ is a schedule. Consider an event $e_w$ occurring in $w_\sigma \cdot e$. Then the event $e_w$ does not occur in $\sigma$, meaning $e_w \xrightarrow{\text{hb}}_{E \cdot w \cdot e} e$ but not $e_w \xrightarrow{\text{hb}}_{E \cdot \sigma} \texttt{last}(\sigma)$. Hence we infer that $e_w$ happens-before $e$ only through events occurring in $\sigma$, i.e., $e_w \xrightarrow{\text{hb}}_{E \cdot w_\sigma \cdot e} e$ infering $w_\sigma \cdot e$ is a schedule.

We intend to apply the I.H. once more for $E \cdot \sigma$ in place of $E$ and $w_\sigma \cdot e$ in place of $w \cdot e$ after establishing $\texttt{UpdSeq}((w_\sigma \cdot e), \textit{SSChar}[E \cdot \sigma])$ is not blocking. A sleep set expression $\psi$ in $\textit{SSChar}[E \cdot \sigma]$ is obtained from two places. Either $\psi$ is returned by the call $\texttt{UpdSeq}(\sigma, \textit{SSChar}[E])$ at line 8. A similar argument in Case 1 establishes that $\psi$ cannot block the $w_\sigma \cdot e$. Or, $\psi$ comes from the return value of calling the function `MkSchedChar()` at line 10. As $\sigma$ is the smallest among the $\sigma_u$s in the $\propto$ order, $\psi$ cannot block the $w_\sigma \cdot e$, thus establishing $\texttt{UpdSeq}((w_\sigma \cdot e), SSChar[E \cdot \sigma])$ is not blocking. The I.H. implies that the call $\texttt{Explore}(E)$ will explore $\texttt{Explore}(E \cdot \sigma \cdot w'')$ for some sequence $w''$ such that (i) $(w_\sigma \cdot e) \simeq e \downarrow^{w''}$, and (ii) $e$ is fresh in $w''$ after $E \cdot \sigma$. These properties, in fact, imply the conclusion of the lemma: Property (i) follows by noting that $e_u \xrightarrow{hb} e$ and that $\sigma \cdot w_\sigma \cdot e \simeq w \cdot e$. Property (ii) follows by the fact that $\sigma$ is the only schedule in $\sigma \cdot w''$ that is not in $w''$, meaning that $e$ is fresh in $\sigma \cdot w''$ after $E$. □

We can now turn to the proof of Theorem 3. It follows a similar structure as the preceding proof:

*Proof of Theorem 3:* We prove Theorem 3 by induction on the maximum length of executions after $E$.

**Base case:** When $E$ is a maximal execution the theorem trivially holds as there is no $w$ to explore after $E$.

**Inductive Step:** Let $e'$ be the first event such that a call of form $\texttt{Explore}(E \cdot e')$ is performed. By the test at line 13, $\texttt{UpdSeq}(\langle e' \rangle, SSChar[E]) \neq block$. There are two cases:

**Case 1:** $e' \sim w$. By analogous reasoning as in Case 1 in the proof of Lemma 3, we show that the I.H. implies that $w \backslash e'$ will be explored after $E \cdot e'$, which proves the theorem in this case.

**Case 2:** $e' \nsim w$. By analogous reasoning as in Case 2 in the proof of Lemma 3, and using Lemma 3, we conclude that the algorithm will generate a schedule $\sigma_u = e_u \downarrow^{u' \cdot e_u}$. For each $e_u$ the algorithm will create one schedule $\sigma_u$. Consider the schedule $\sigma$ among $\sigma_u$s that the algorithm chose to explore last after $E$. It gives rise to a call $\texttt{Explore}(E \cdot \sigma)$. We can then again apply the I.H. to establish the conclusion of the theorem.    □

The following theorem, which is similar of Theorem 3.2 of Godefroid *et al.* [22] establishes optimality of POP.

**Theorem 4.** *POP never explores two maximal execution sequences which are equivalent.*

*Proof.* Assume that $E_1$ and $E_2$ are two equivalent maximal execution sequences, which are explored by POP. Let $E$ be their longest common prefix such that $E_1 = E \cdot e_1 \cdot w_1$ and $E_2 = E \cdot e_2 \cdot w_2$. By the rules of race reversal, the last event of $E$ is the head of a schedule, and the schedule ends with $E$. Thus, the algorithm must have performed the first recursive call of form $\texttt{Explore}(E \cdot e)$ at line 15 during exploration of $E$. If $e$ is equal to $e_1$ (or $e_2$), then by construction, $e_2$ (or $e_1$) must be in a schedule that is not compatible with $e_1$ (or $e_2$), implying $E_1$ and $E_2$ are not equivalent. Otherwise, neither $e_1$ nor $e_2$ equals to $e$, meaning $e_1$ and $e_2$ must be in different schedules ($\sigma_1$ and $\sigma_2$) that race with $e$. These schedules are mutually incompatible, except if their heads are read on the same variable. But then inequivalence of $e_1 \cdot w_1$ and $e_2 \cdot w_2$ is achieved by the sleep set mechanism, which is proven correct by Lemma 2. □

**Theorem 5.** *Algorithm 1 is never blocked (at line 13) unless $E$ is a maximal execution.*

*Proof.* The only reason for which Algorithm 1 can get blocked at line 13 is that the test fails, i.e., for each event $e$ in $\texttt{enabled}\,(E)$, the call $\texttt{UpdSeq}(e, \mathit{SSChar}[E])$ is blocking, which cannot happen for the following reason. Consider an arbitrary sleep set expression $\psi$ from $\mathit{SSChar}[E]$, which was added (line 11) because of the reversal of a race involving the first event, a write event $e'$, and the second event, a read event $e''$. Note that the write event $e'$ is enabled in every continuation after the race reversal, and $\texttt{UpdSeq}(e', \mathit{SSChar}[E])$ (line 11 of Algorithm 2) is not blocking and removes $\psi$ from $\mathit{SSChar}[E]$. Hence, the contradiction proves the theorem. □

**Lemma 1** *The number of nested recursive calls to* $\texttt{Explore}$ *at line 12 is at most $\mathcal{O}(n^2)$, where $n$ is the length of the longest execution of the program.*

Note that in this lemma, we do not count the calls at line line 15, since they are considered as normal exploration of some execution. Only the calls at line 12 start the exploration of a new execution.

*Proof.* The algorithm explores executions by calling $\texttt{Explore}$ recursively with the explored execution as an argument. For a recursive call $\texttt{Explore}\,(E \cdot \sigma)$, assign a *level* denoting the length of the prefix $E$ and a *degree* denoting the number of schedule events in the argument execution $E \cdot \sigma$. The freshness condition of the second event of a parsimonious race and the schedule construction mechanism imply that all the schedules occurring in $\texttt{Explore}\,(E \cdot \sigma)$ also occur in the next recursive call $\texttt{Explore}\,(E' \cdot \sigma')$ inside $\texttt{Explore}\,(E \cdot \sigma)$, meaning the degree either stays same or increases in the next recursive call. The degree stays the same when the schedule $\sigma'$ is produced by a parsimonious race where the first racing event occurs in $E$, meaning $E'$ will be shorter than $E$. From the above fact, we can say that with each recursive call, either the degree increases or the level decreases. The degree can be from 0 to $n-1$, meaning it can increase $n$ times. The level can be between 0 and $n-2$, meaning it can decrease $n-1$ times. Hence, there are at most $\mathcal{O}(n^2)$ recursive calls. □

For the proof of Theorem 2, we need the following lemma, which bounds the size of sleep set characterizations.

**Lemma 4.** *In Algorithm 1, each sleep set characterization $\mathit{SSChar}[E]$ for an explored execution $E$ contains at most a quadratic number of sleep set expressions, in the length of the longest program execution.*

*Proof.* A sleep set characterization consists of a set of sleep set expressions. Each contained sleep set expression $\psi$ is the result of propagating a schedule expression $\varphi$ through a part of $E$. Each schedule expression $\varphi$ has an innermost expression of form $P \rhd x$, generated at some race reversal; thereafter, this innermost expression has been enclosed in an expression of form $P'[\varphi] \rhd x'$ whenever an inherited schedule expression has been formed. For an execution sequence $E$, now consider the set of schedule expressions that occur in some schedule characterization associated with a schedule in $E$. We claim that each such innermost schedule expression is derived from a race reversal that generated an enclosing call to `Explore` (at line 12), and that each such innermost schedule expression occurs at most once in the schedule characterizations in $E$. This claim can be checked by induction over the sequence of enclosing calls to `Explore` in which the current exploration of $E$ is situated: the inductive step inspects that the claim is maintained. From this claim and Lemma 1, it follows that each sleep set expression can have at most a qua dratic number of sleep set expressions. Each of them has the size at most linear in the length of the longest program execution. □

**Theorem 2.** *Algorithm 1 needs space $\mathcal{O}(n^6)$, where $n$ is the length of the longest execution of the analyzed program.*

*Proof.* By Lemma 1, at any point in time the algorithm needs to maintain information for at most a quadratic number of executions. For each such execution $E$, the most space-consuming information consists of a sleep set characterization for each prefix of $E$. By Lemma 4, each such characterization consists of at most a quadratic number of sleep set expressions, each of which is at most linear in the length of the longest execution. Since sleep set characterizations are associated with each prefix of an execution, the worst-case memory consumption is at most $\mathcal{O}(n^6)$, where $n$ is the length of the longest program execution. □

## B  An example with potentially exponential size sleep sets

In this section, we show a program, which is parametric on the number of threads spawned, that leads to storing an exponential number of schedules in POP.

The code of the program (Fig. 3(left)), consists of a thread $p$ doing a write on x, a thread $q$ that spawns $n$ threads, $q_1, \ldots, q_n$, each doing a write on y, followed by thread $q$ performing a read from x to a local register $a$ after all the children threads have been joined. Optimal DPOR explores $2n!$ executions, shown in the exploration tree in the right part of Fig. 3. Optimal DPOR's depth-first exploration strategy first explores $n!$ executions $E_1, \cdots E_{n!}$ as a result of the reversal of races involving $n$ write events on y ($n!$ permutations of $n$ write events). In each execution $E_i$, the algorithm reverses the race (in red arrow) between the events `x = 1` and `a = x`, and stores the produced schedule $\sigma_i$. Eventually, Optimal DPOR stores $n!$ schedules. In contrast, POP's eager race reversal strategy eliminates this problem by exploring each $\sigma_i$ immediately after production and redundancy check.

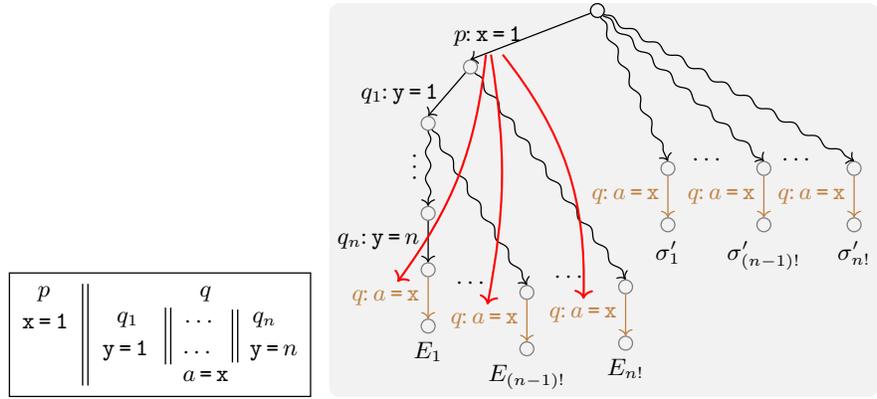

**Fig. 3.** The exp-mem3 program (left) and its incomplete exploration tree (right).

However, there is also another problem: there are an exponential number of schedules in the sleep set $sleep[\langle\rangle]$. This happens because after finishing the exploration, each $\sigma_i$ ends up in the sleep set $sleep[\langle\rangle]$. When the last schedule $\sigma_{last}$ is produced, $sleep[\langle\rangle]$ contains all the $(n!-1)$ previously explored $\sigma_i$'s. POP's parsimonious sleep set characterization (Section 4.3) solves this problem by representing sleep sets by polynomial size expressions. POP not only solves the problem with exponential space, but also makes the redundancy check much faster in this case, as it does not have to compare against an exponential number of schedules in the sleep set. The experimental results we present in the next section (benchmark exp-mem3 of Table 1) show exactly that.